\documentclass[prd,twocolumn,showpacs,groupedaddress,superscriptaddress,amsmath,amssymb]{revtex4-2}

\usepackage{hhline}
\usepackage{slashed}
\usepackage{graphicx}% Include figure files
\usepackage{graphics}
\usepackage{dcolumn}% Align table columns on decimal point
\usepackage{bm}% bold math
\usepackage{amssymb}
\usepackage{enumerate}
\usepackage{lineno}
\usepackage{multirow}
\usepackage{float}
\begin{document}

%\title{Sensitivity of NA64$\mu$ experiment at the CERN SPS to search for a new light $Z'$ boson}
\title{Prospects in the search for a new light $Z'$ boson\\
with the NA64$\mu$ experiment at the CERN SPS}

\author{H. Sieber}
\affiliation{ETH Zürich, Institute for Particle Physics and Astrophysics, CH-8093 Zürich, Switzerland}
\author{D. Banerjee}
\affiliation{CERN, European Organization for Nuclear Research, CH-1211 Geneva, Switzerland}
\author{P. Crivelli}
\affiliation{ETH Zürich, Institute for Particle Physics and Astrophysics, CH-8093 Zürich, Switzerland}
\author{E. Depero}
\affiliation{ETH Zürich, Institute for Particle Physics and Astrophysics, CH-8093 Zürich, Switzerland}
\author{S. N. Gninenko}
\affiliation{Institute for Nuclear Research of the Russian Academy of Sciences, 117312 Moscow, Russia}
\author{D. V. Kirpichnikov}
\affiliation{Institute for Nuclear Research of the Russian Academy of Sciences, 117312 Moscow, Russia}
\author{M. M. Kirsanov}
\affiliation{Institute for Nuclear Research of the Russian Academy of Sciences, 117312 Moscow, Russia}
\author{V. Poliakov}
\affiliation{State Scientific Center of the Russian Federation Institute for High Energy Physics of National Research Center ’Kurchatov Institute’ (IHEP), 142281 Protvino, Russia}
\author{L. Molina Bueno}
\email[corresponding author: ]{laura.molina.bueno@cern.ch}
\affiliation{ETH Zürich, Institute for Particle Physics and Astrophysics, CH-8093 Zürich, Switzerland}
\affiliation{CSIC - Universitat de València, Instituto de Física Corpuscular (IFIC), E-46980 Paterna, Spain}

\date{\today}

\begin{abstract}
  %Dark sectors weakly coupled to the second and third lepton generations through $L_{\mu}-L_{\tau}$ current are an interesting framework to explain the origin of Dark Matter. 
  A light $Z'$ vector boson coupled to the second and third lepton generations through $L_{\mu}-L_{\tau}$ current with mass below 200 MeV provides a very viable explanation in terms of new physics to the recently confirmed $(g-2)_\mu$ anomaly. This boson can be produced in the bremsstrahlung reaction $\mu N \rightarrow \mu N Z'$ after a high energy muon beam collides with a target. NA64$\mu$ is a fixed-target experiment using a 160 GeV muon beam from the CERN Super Proton Synchrotron accelerator looking for the $Z'$ production and its subsequent decays, $Z'\rightarrow invisible$. 
  %In 2021, a pilot run has been approved and the first physics run is expected in 2022. 
  In this paper, we present the study of the NA64$\mu$ sensitivity to search for such a $Z'$.  This includes a realistic beam simulation, the detailed detectors description and a discussion about the main potential background sources. A pilot run is scheduled in order to validate the simulation results. If those are confirmed, NA64$\mu$ will be able to explore all the remaining phase space which could provide an explanation for the $g-2$
  muon anomaly.
\end{abstract}

\maketitle

\section{Introduction} \label{sec:introduction}

The recently confirmed  4.2$\sigma$ deviation of the muon magnetic moment \cite{Abi:2021gix} with respect to its Standard Model (SM) prediction \cite{Aoyama:2020ynm,Aoyama:2012wk,Aoyama:2019ryr,Czarnecki:2002nt,Gnendiger:2013pva,Davier:2017zfy,Keshavarzi:2018mgv,Colangelo:2018mtw,Hoferichter:2019gzf,Davier:2019can,Keshavarzi:2019abf,Kurz:2014wya,Melnikov:2003xd,Masjuan:2017tvw,Colangelo:2017fiz,Hoferichter:2018kwz,Gerardin:2019vio,Bijnens:2019ghy,Colangelo:2019uex,Blum:2019ugy,Colangelo:2014qya} might be an indication of physics beyond the SM:

\begin{equation}
  \Delta a_{\mu} \equiv a_{\mu}(\text{exp}) - a_{\mu}(\text{th}) =(251 \pm 59)\cdot 10^{-11}    
\end{equation}
Interactions between muons and new physics (NP) sectors have been suggested in many models \cite{LINDNER20181,Stockinger:2013rna,doi:10.1146/annurev-nucl-031312-120340}. In particular, models with $U(1)$ gauge extension to the SM are well motivated since they are anomaly-free and provide an explanation to the $(g-2)_\mu$ anomaly through a loop contribution to the muon vertex function \cite{Gninenko:2001hx, Gninenko:2014pea, Chen:2017awl, Gninenko:2018tlp, Kirpichnikov:2020tcf, Amaral:2021rzw}. In the $L_\mu-L_\tau$ model, with SM gauge extension $SU(3)_c\otimes SU(2)_{L}\otimes U(1)_{Y}\otimes U(1)_{L_\mu-L_\tau}$ \cite{Foot:1990mn,PhysRevD.43.R22,PhysRevD.44.2118,Gninenko:2018tlp}, the massive gauge vector boson $Z'$ acquires its mass through symmetry breaking, $m_{Z'}\leq\mathcal{O}(1\ \text{GeV})$, and interacts to the second and third generations of leptons through:
\begin{equation}
  \label{eq:lagrangian}
  \mathcal{L}=g'(\bar{\mu}\gamma_\alpha\mu+\bar{\nu}_\mu\gamma_\alpha\nu_\mu-\bar{\tau}\gamma_\alpha\tau+\bar{\nu}_\tau\gamma_\alpha\nu_\tau)Z'^{\alpha},
\end{equation}
where $Z'^\alpha$ is the leptophilic boson field and $g'$ is its coupling to SM leptons. Within this model, the $Z'$ contribution to the muon vertex function, and thus the muon $(g-2)_\mu$, is estimated at one loop to be \cite{Baek_2009}:
\begin{equation}
  \label{eq:one_loop}
  \Delta a_\mu^{Z'}=\frac{g'^2}{4\pi^2}\int_{0}^{1}dx\ \frac{x^{2}(1-x)}{x^2+(1-x)m_{Z'}^{2}/m_\mu^{2}}
\end{equation}
The $Z'$ vector boson decays invisibly to SM neutrinos in the case $m_{Z'}<2m_\mu$, with decay width:
\begin{equation}
  \label{eq:invisible_decay}
  \Gamma(Z'\rightarrow \bar{\nu}_f\nu_f)=\frac{\alpha_\mu m_{Z'}}{3},
\end{equation}
where $f=\mu,\ \tau$ and $\alpha_\mu=g'^2/4\pi$. For larger $Z'$ masses, namely $m_{Z'}\geq2m_\mu$, the gauge boson also decays visibly to one of the charged component of the $SU(2)_L$ leptons doublets, $L_\mu$, $L_\tau$, with partial decay width:
\begin{equation}
  \label{eq:visible_decay}
  \Gamma(Z'\rightarrow \bar{f}f)=\frac{\alpha_\mu m_{Z'}}{3}\cdot\left(1+\frac{2m_f^2}{m_{Z'}^2}\right)\cdot\sqrt{1-\frac{4m_f^2}{m_{Z'}^2}}
\end{equation}
It is worth noting that adding to the minimal $U(1)_{L_\mu-L_\tau}$ gauge extension of the SM a dark current interaction of the type $\mathcal{L}\supset Z'^\alpha J_\alpha^{DM}$ makes it possible to also probe light thermal dark matter (LTDM) and the Dark Matter (DM) relic abundance (see e.g. \cite{Banerjee:2019na64mu, Kahn:2018cqs}). \\
The $Z'$ vector boson can be produced through muon bremsstrahlung $\mu N\rightarrow\mu NZ'$ after a high energy muon beam impinges on an active target. Within this context, the NA64$\mu$ experiment \cite{Banerjee:2019na64mu} has been designed to search for the $Z’$ production and subsequent invisible decay using the 160 GeV M2 beam-line at the CERN SPS accelerator \cite{Doble:1994np}. Detailed computations of the differential and total cross-sections for this process have been recently performed in \cite{kirpichnikov2021probing}. Another experiment, M$^3$, with a similar working principle, has been proposed at Fermilab \cite{Kahn:2018cqs}. A pilot run of NA64$\mu$ experiment is planned in the end of 2021 to study the feasibility of the technique. In this paper we discuss the experiment prospects in terms of the main background sources and the expected trigger rate. We also study the projected sensitivities for future physics run to allow to probe the region of parameter space suggested by the $(g-2)_\mu$ anomaly in the context of current and future searches.

\section{The method of  search}\label{sec:experiment}
\begin{figure*}[]
  \centering
  \includegraphics[width=0.95\textwidth]{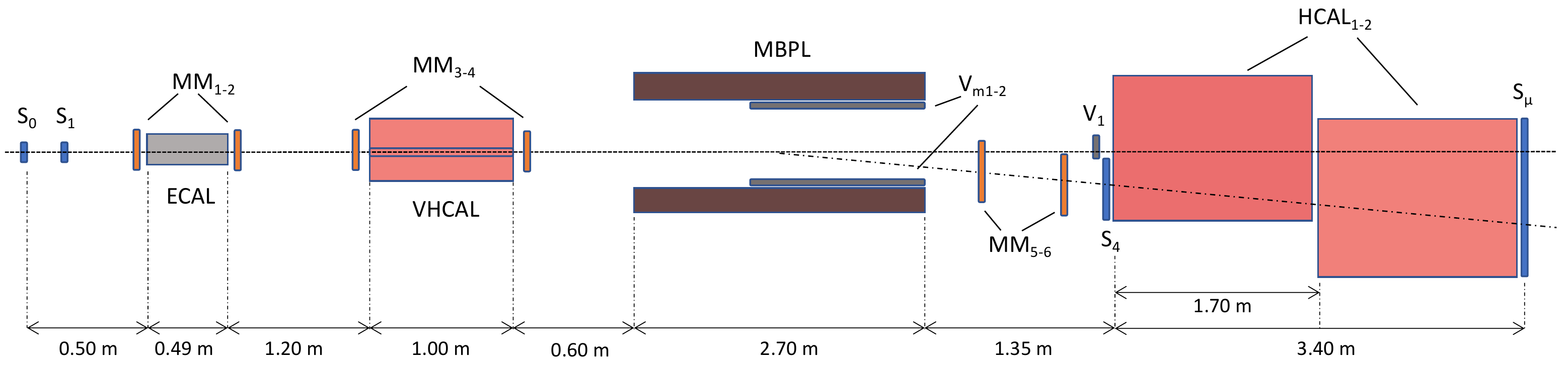}
  \caption{\label{fig:experimental-set-up}Schematic view of the 2021 muon pilot run experimental set-up to search for $Z'\rightarrow invisible$ production from 160 GeV/c muon bremsstrahlung.}
\end{figure*}
\begin{figure}[H]
  \centering
  \includegraphics[width=0.45\textwidth]{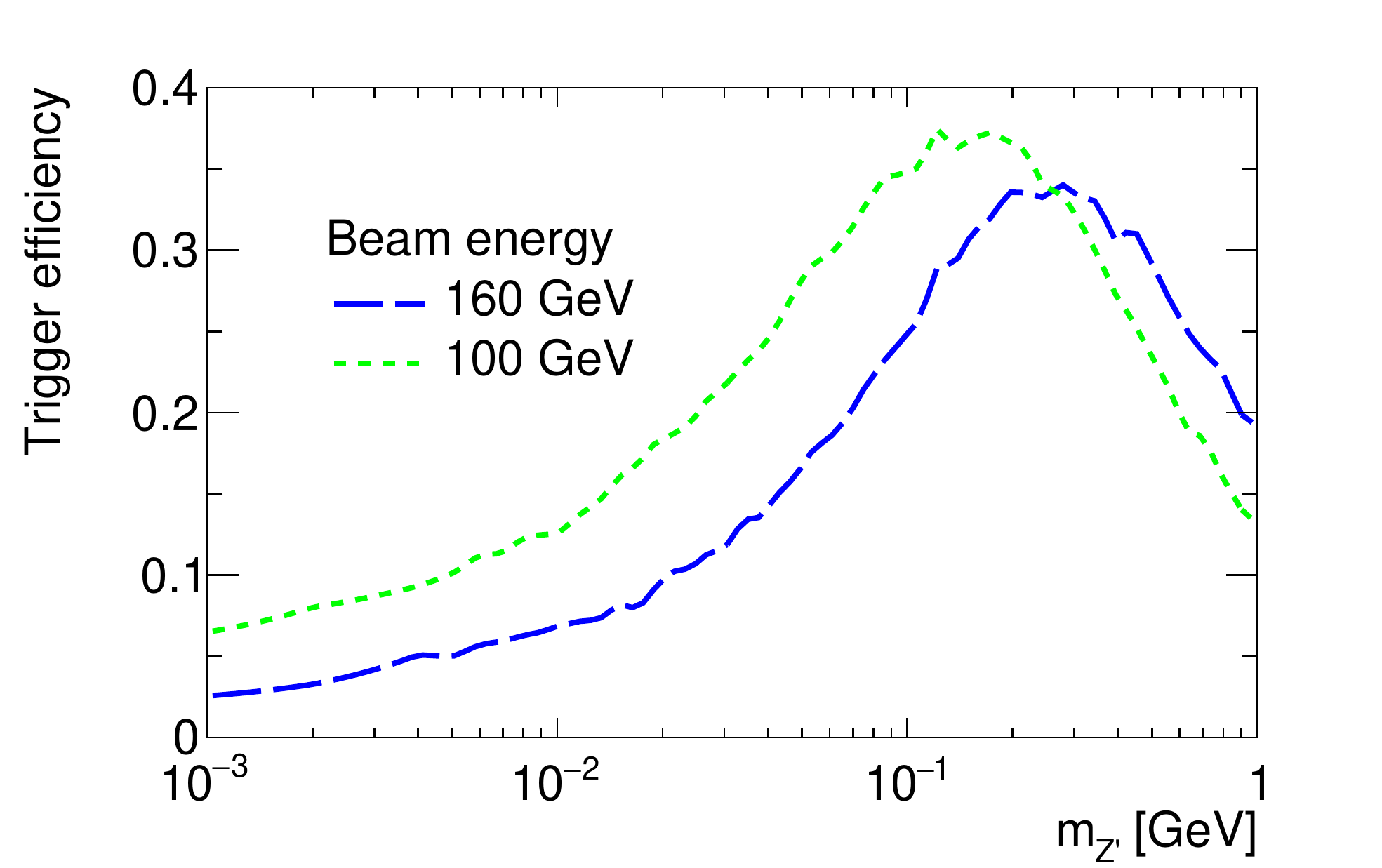}
  \caption{\label{fig:trigEff}Trigger efficiency as a function of the $Z'$ vector boson mass. The beam nominal energy is both (blue) 160 GeV and (green) 100 GeV, for which the scattered muon deflection is larger.}
\end{figure}

The NA64$\mu$ experiment \cite{Banerjee:2019na64mu}, is a complementary experiment to NA64$e$ \cite{Andreas:2013lya,Gninenko:2013rka} aiming to look for dark sectors weakly coupled to muons. The experiment is foreseen in two phases. Its first phase physics goal is to search for invisible decays of the $Z'$ boson, produced in the muon scattering process $\mu^{-}N\rightarrow\mu^{-}NZ'$. Additionally, similarly to the electron mode, the experiment also explores the production of dark photons, $A'$, through the bremsstrahlung $\mu^{-}N\rightarrow\mu^{-}NA'$, allowing to enlarge the parameter space of interest towards large masses \cite{Gninenko:2019qiv}. NA64$\mu$ can also probe scalar, axion-like particles (ALPs), millicharged particles \cite{Gninenko:2018ter} and could also be used to test lepton Flavour Violation (LVF) in $\mu N\rightarrow \tau X$ conversion in flight \cite{Gninenko:2018num}. A second phase of the experiment will be devoted to explore these processes \cite{Banerjee:2019na64mu}. \\
The experimental set-up for the feasibility studies to look for a light $Z'$ boson is sketched in Fig. \ref{fig:experimental-set-up}. The experiment will use the high-energy muon beam M2 at the CERN SPS \cite{Doble:1994np, Bernhard:2019jqz} with momentum $\simeq160$ GeV/c produced by a 450 GeV/c primary proton beam (intensity $10^{12}$-$10^{13}$ protons/spill). Within this context, muons are dumped against an active target, which is a 40 radiation lengths (40$X_0$) shashlik-like (lead-scintillator) electromagnetic calorimeter (ECAL), with a $6\times5$ cell matrix structure. While the scattered muon carries away a fraction $E_\mu^{'}=fE_\mu$ of the primary muon energy $E_\mu$, the other fraction of the energy, $(1-f)E_\mu$, is carried away by the bremsstrahlung dark boson $Z'$ and its decay products resulting in missing energy $E_\text{miss}=E_\mu-E_\mu^{'}$. The sub-detectors downstream the target include, in particular, a 5 interaction lengths (5$\lambda_I$) copper-scintillator veto calorimeter (VHCAL) segmented with a $4\times4$ matrix of cells and a hole in the middle, to veto charged secondaries produced by upstream muon nuclear interactions. Then, a series of two large $120\times60$ cm$^2$ ($6\times3$ matrix) hadronic calorimeter (HCAL) modules, with $7.5\lambda_I$ steel-scintillator, ensure maximal hermeticity. The experiment will use two magnet spectrometers in order to reconstruct the incoming and outgoing muon momentum. The initial muon beam momentum will be measured by the existing Beam Momentum stations (BMS) from the COMPASS experiment \cite{COMPASS:2007rjf}. A set of Micro-MEsh Gaseous Structure (Micromegas) tracking detectors will be located next to the stations to have a second measurement of the incoming momentum. The scattered muon momentum is reconstructed through a second magnetic spectrometer (a single dipole magnet with 1.4 T$\cdot$m, MS$_2$) with a set of six Micromegas tracking detectors. \\
A signal event, i.e. the production of a $Z'$-boson, is defined as a scattered muon after the target with momentum about half of the beam nominal energy ($E_{\mu}'\lesssim0.5E_0\simeq80$ GeV) (see Fig. \ref{fig:experimental-set-up}). The muon missing momentum will be used to keep the trigger rate at the required level. Using the deflection of the scattered muon in the transverse plane, a simplified trigger system has been considered for the initial phase to run the experiment at the low beam intensity of $10^7 $ $\mu/$spill. This trigger option is based on the selection of the well-defined primary muon beam within the small lateral size, divergency, and momentum spread by using small size beam defining counters, the BMS station and the trackers next to it. The initial muon is tagged with a set of three scintillator counters, two of them before the target ($S_0$ and $S_1$), and one ensuring the presence of the muon at the end of the set-up ($S_\mu$). %To maximise the signal efficiency detection and keeping the rate from multiple scattering from the primary beam as low as possible, an additional counter in front of HCAL, $S_4$, shifted from the beam axis is used to trigger on the deflected muons.
An additional counter in front of the HCAL, $S_4$, shifted from the beam axis guarantees that only muons with enough deflection hit this counter. The signal efficiency for different $Z'$ masses at this trigger level is illustrated in Fig. \ref{fig:trigEff}. A set of veto counters, before the HCAL modules ($V_1$), and within MS$_2$ ($V_{\text{m}_{1,2}}$), are used to further veto undeflected beam muons and veto charged secondaries produced in upstream interactions. The trigger rate after this selection is 0.1\% of the primary beam intensity. \\
To further suppress background coming from SM events, a set of selection criteria (cuts) is applied as follows, (i) an initial beam momentum reconstructed in the energy window [140, 180] GeV, (ii) a single track in the tracking detectors with reconstructed momentum smaller than half of the beam energy (iii) no energy deposit in the VHCAL, (iv) no energy deposit in the HCAL modules (i.e. compatible with the one of a minimuon ionizing particle (MIP)).
\section{Simulation framework}
Detailed Monte Carlo (MC) simulations are performed using the \texttt{GEANT4} toolkit \cite{Agostinelli:2002hh} and the Geant4 compatible DM package \texttt{DMG4} \cite{Celentano:2021cna} aiming at realistically reproducing the beam-line, detectors and physics, as well as estimating the background and signal topology within the set-up. \\
\subsection{The beam profile at M2 location}
\begin{figure*}[]
  \centering
  \includegraphics[width=0.45\textwidth]{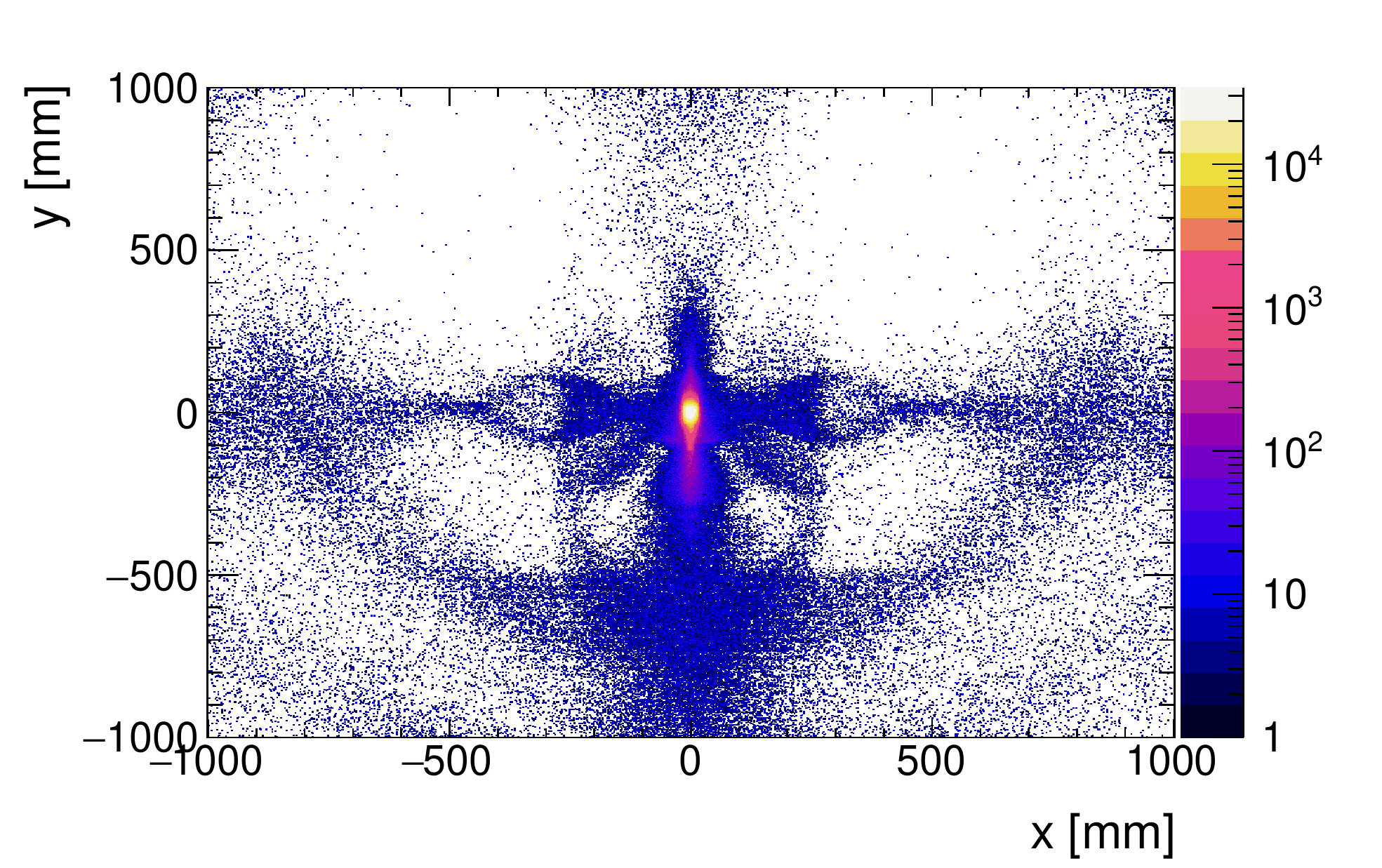}
  \hspace{5mm}
  \includegraphics[width=0.45\textwidth]{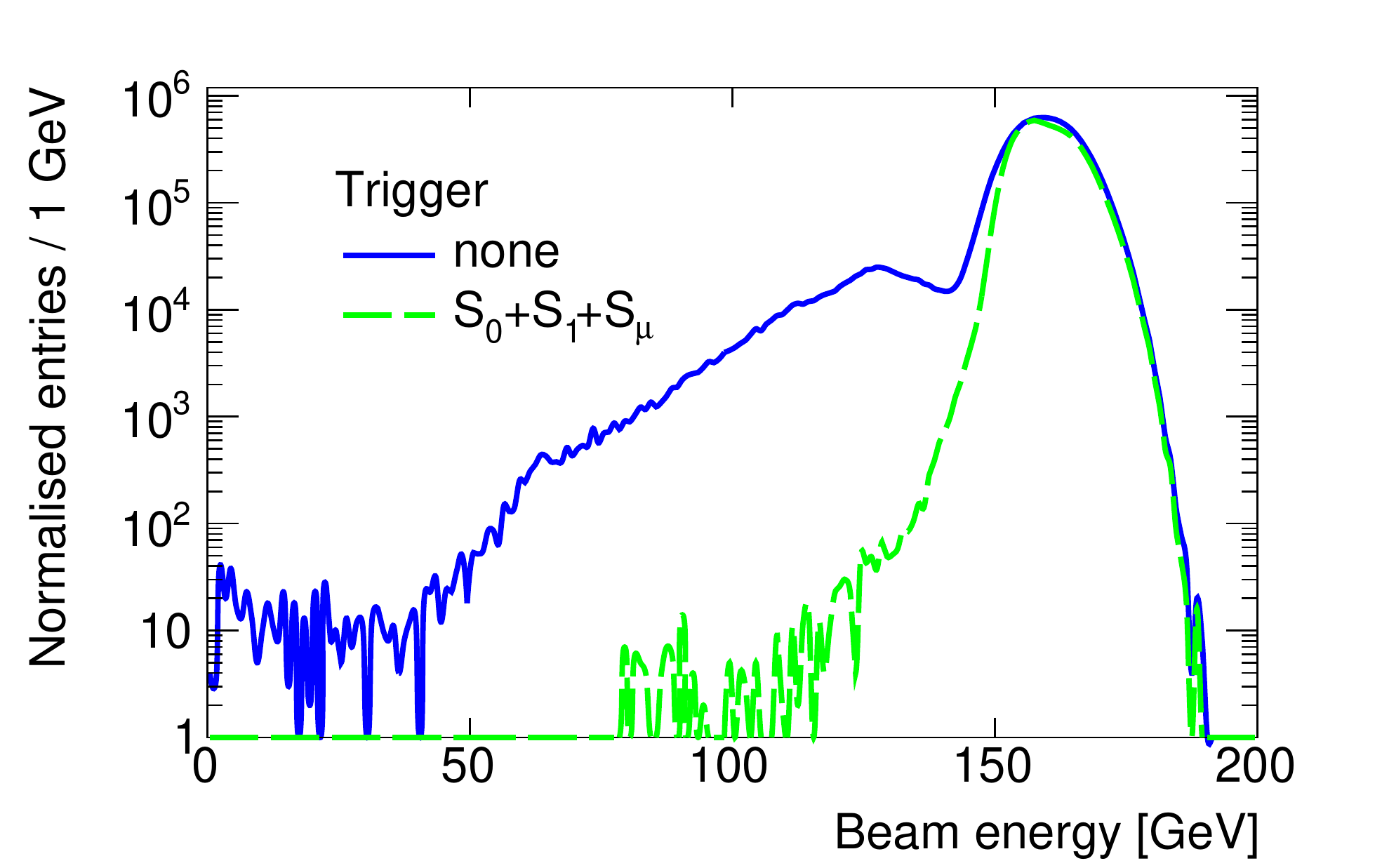}
  \caption{\label{fig:beamProf}Beam profiles at the entrance of the NA64$\mu$ set-up as obtained through the \texttt{TRANSPORT} \cite{Brown1980}, \texttt{TURTLE} \cite{Turtle1974} and \texttt{HALO} \cite{Iselin1974} software  \cite{Bernhard:2019jqz} for (\emph{Left}) beam spatial distribution and (\emph{Right}) beam energy spectrum with no trigger (blue) and (red) trigger $S_{0}+S_{1}+S_{\mu}$.}
\end{figure*}

Because of the importance of an accurate knowledge of the initial muon momentum and beam spatial distribution for the trigger criteria, the M2 beam-line optics is fully simulated using the \texttt{TRANSPORT} \cite{Brown1980}, \texttt{TURTLE} \cite{Turtle1974} and \texttt{HALO} \cite{Iselin1974} software  \cite{Bernhard:2019jqz} and made compatible to \texttt{GEANT4} through the \texttt{HepMC} software \cite{Dobbs:684090}. This yields realistic beam profiles as shown in Fig. \ref{fig:beamProf}. Simulations reproduce the large contribution of low-energy halo muons around the beam spot (about 20\% of the full beam intensity \cite{Banerjee:2019na64mu}) that are populating the low-energy tail of the beam energy distribution. Such muons can be efficiently removed using the beam-defining counters $S_0$ and $S_1$ , leaving 78\% of the full beam intensity, with muons energy $\geq100$ GeV, with beam spot size $\sigma_x\sim0.9$ cm and $\sigma_y\sim1.9$ cm.

\subsection{Signal}
\begin{figure}[H]
  \centering
  \includegraphics[width=0.45\textwidth]{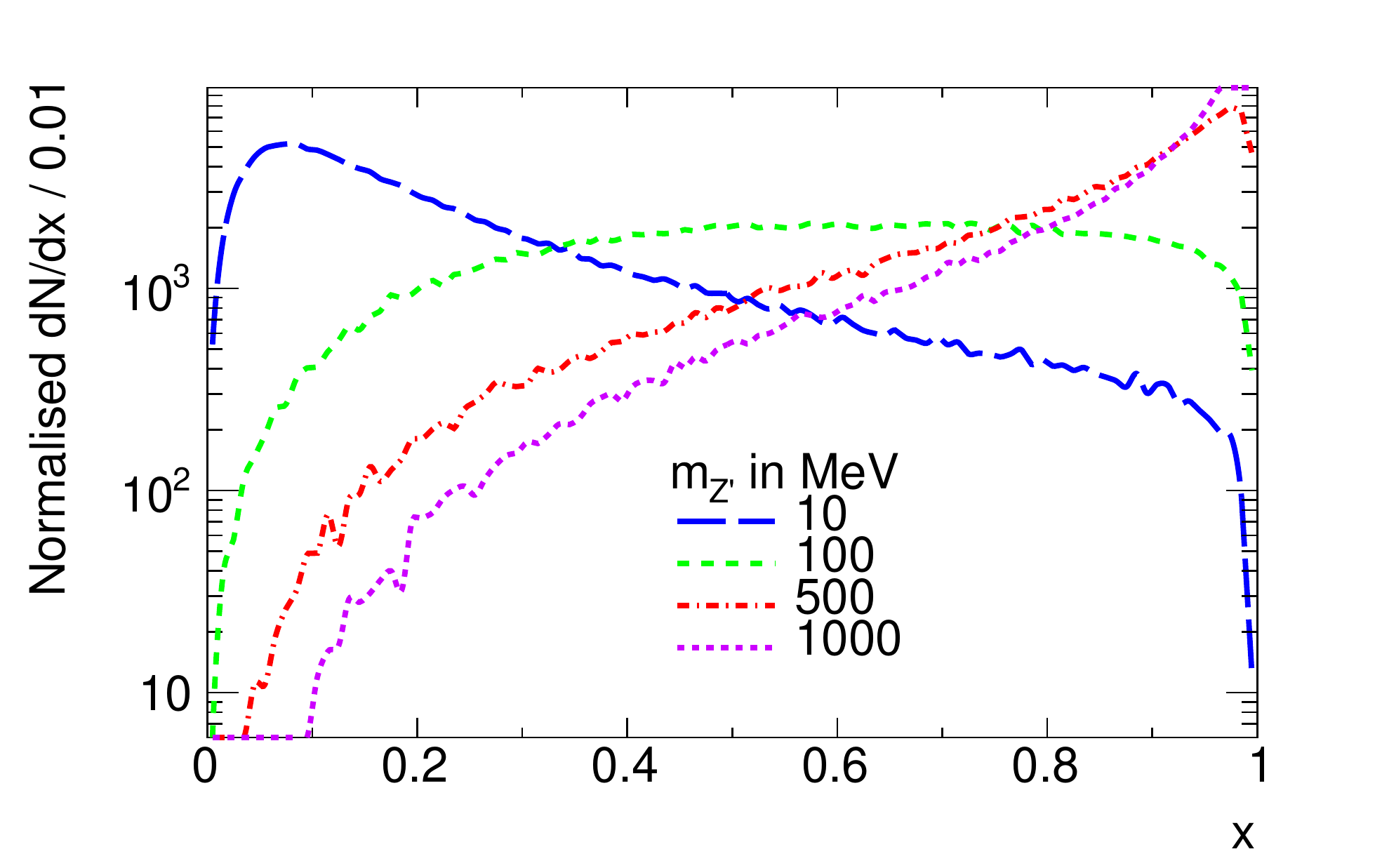}
  \caption{\label{fig:Zspectra}Normalised energy spectra of $Z'$ vector boson for different masses obtained from \texttt{GEANT4} \cite{Agostinelli:2002hh} using the \texttt{DMG4} \cite{Celentano:2021cna} package. The mixing strength is $\epsilon=g'/\sqrt{4\pi\alpha}=10^{-4}$, where $\alpha=1/137$. See \cite{kirpichnikov2021probing} for more details.}
\end{figure}

To estimate both the signal yield and signal topology, and thus the choice of adequate selection criteria, the $Z'$ vector boson is simulated using the fully \texttt{GEANT4}-compatible \texttt{DMG4} package \cite{Celentano:2021cna}. Dark matter observables, such as total cross-section production and differential cross-sections are correspondingly implemented according to the Weisz\"{a}cker-Williams (WW) phase-space approximations as discussed in \cite{kirpichnikov2021probing}. %After a $Z'$ production, 
The program selects an event where the $Z'$ production should occur according to the total cross section, and then both its fractional energy, $x$, and emission angle, $\theta$, are accurately sampled using a Von-Neumann accept-reject sampling algorithm. The typical energy spectra of a $Z'$-strahlung vector boson is shown in Fig. \ref{fig:Zspectra} for the mass range 10 MeV to 1 GeV.

\subsection{Background}
\begin{figure}[h]
  \centering
  \includegraphics[width=0.45\textwidth]{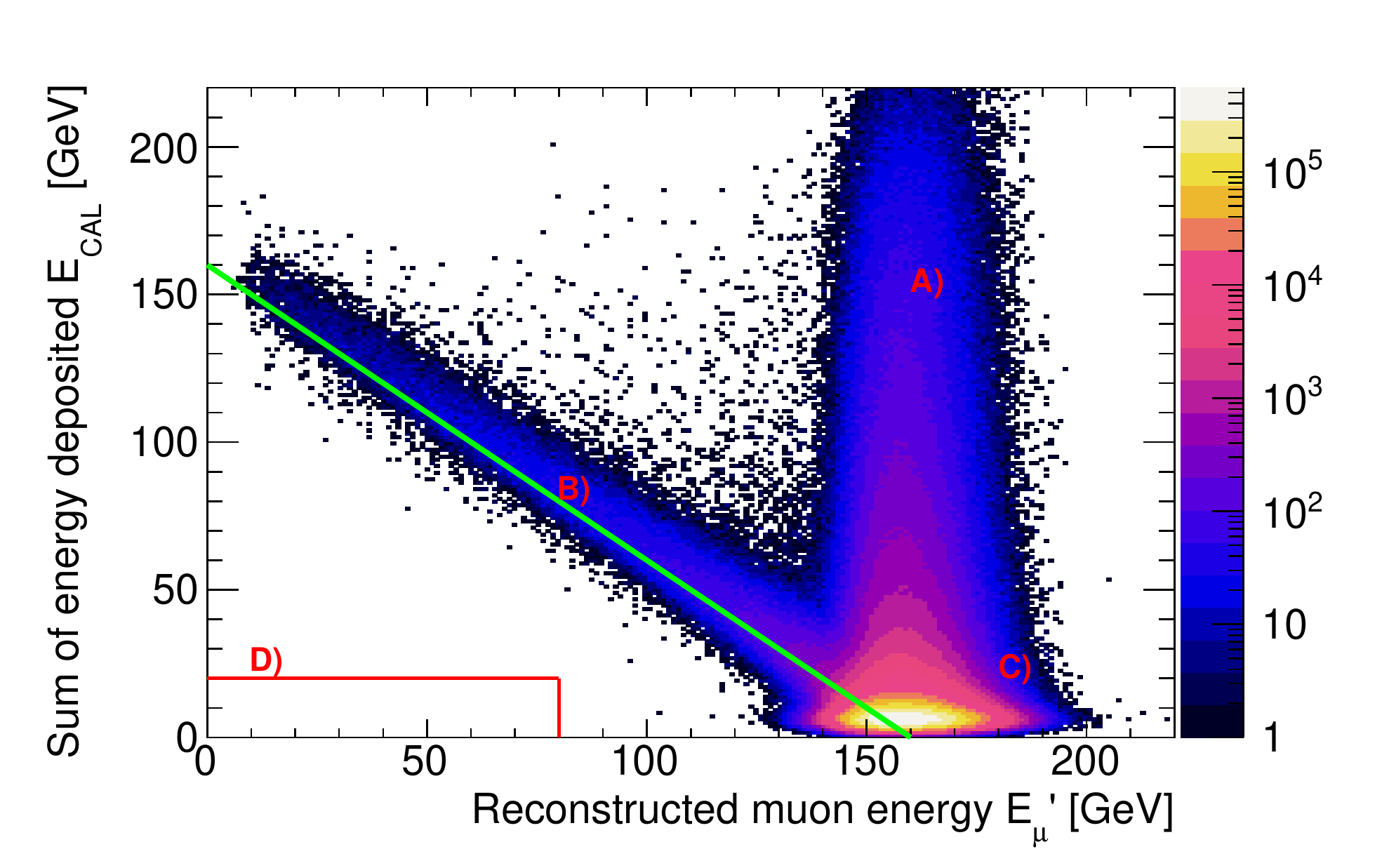}
  \hspace{5mm}
  \includegraphics[width=0.45\textwidth]{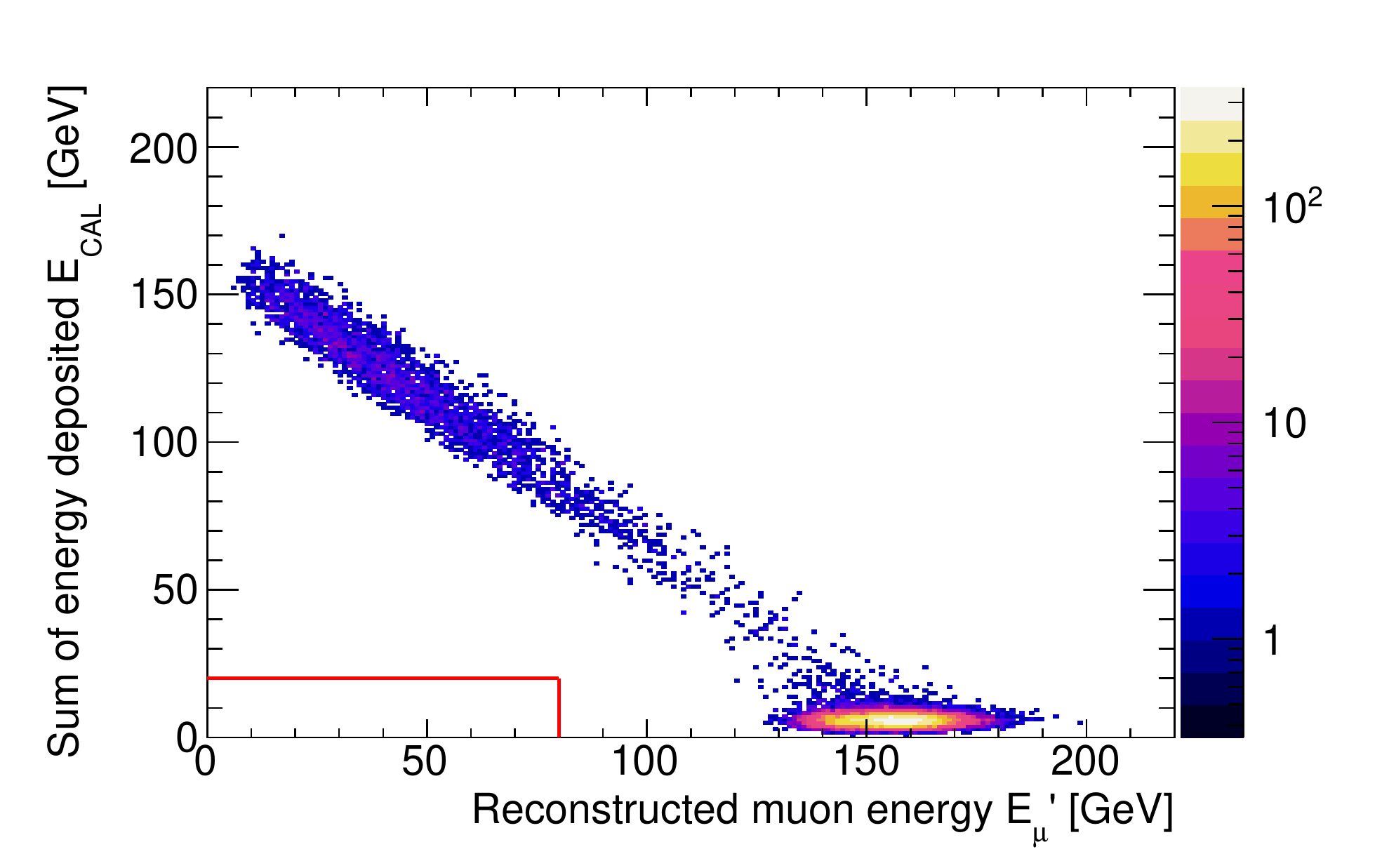}
  \caption{\label{fig:hermeticity}Hermeticity plane defined by the reconstructed muon energy after the ECAL and the total energy deposit in the calorimeters, $(E_\mu';\ E_\text{CAL}=E_\text{ECAL}+E_\text{VHCAL}+E_\text{HCAL})$, for SM muons with a single reconstructed track (one hit per Micromegas) assuming (\emph{Top}) trigger in counters $S_0$, $S_1$ and $S_\mu$ and  (\emph{Bottom}) trigger in all counters, no energy deposit in the vetos, VHCAL and HCAL. The total number of simulated muons corresponds to $N_\text{MOT}=10^{8}$. Whereas region $A$, $B$ and $C$ correspond to regions compatible with SM-related energy desposit, region $D$ defines the interesting parameter space compatible with $Z'$-strahlung events (see text).}
\end{figure}

Missing energy/momentum experiments such as NA64$\mu$ rely not only on a robust detector hermeticity in order to avoid events with large missing energy appearing because some particles escaped detection due to acceptance, but also on a precise momentum reconstruction. Many processes, such as hard muon nuclear interactions in the ECAL, hadron admixture in the M2 beam-line, or mismatch in momentum reconstruction, can affect the likelihood to truly observe a $Z'$-strahlung process. In the following paragraphs, the full study of those main background contributions is covered, being carried out through detailed MC simulations. \\ 
Muons usually behave as MIP, thus most of them traverse the whole set-up with nominal momentum $\sim160$ GeV/c, with small energy deposit in the calorimeters ($E_\text{ECAL}\sim0.5$ GeV, $E_\text{HCAL}\sim2.3$ GeV). On the opposite, scattered muons accompanied by $Z'$ emission are identified with energy $E_\mu'\leq0.5E_0\simeq80$ GeV. Accurate momentum reconstruction thus allows to discriminate between possible signal candidates and SM muon events. The muon momentum is reconstructed both upstream the target and downstream the ECAL through a series of magnetic spectrometers. To account for multiple scattering (material) effects as well as tracking detectors resolution, the precision on momentum reconstruction was estimated using the Kalman-filter-based \texttt{GenFit} toolkit \cite{Rauch:2014wta} to be $\Delta p/p\sim3\%$. The mis-measurement, and thus mis-identification between $Z'$-strahlung and SM muons, is extracted from the exponential tails of the momentum residuals distribution. Through extrapolation, the probability of a 160 GeV SM muon to be reconstructed with momentum $\leq80$ GeV is $\lesssim10^{-12}$ per MOT. This result is obtained assuming that in particular selection criterium (ii) holds, i.e. a single hit per tracking detectors. In the case of more than one hit per tracker, the MM detector inefficiency should be taken into account. An example of such a physics process are highly energetic secondaries produced by muons in the tracker material through ionisation, $\mu e\rightarrow\mu e$, accompanied by a poorly detected parent muon, and thus yielding a reconstructed momentum with energy $<100$ GeV. The probability for such an event to happen is estimated from the full sample of simulated muons, taking into account the values for Micromegas trackers inefficiency ($\sim0.02$) extracted by previous NA64$e$ run data, and assuming that in the second tracker downstream MS2 there is more than one hit. From the simulations, a conservative value of $\leq10^{-11}$ per MOT is obtained, and can be further reduced by placing additional $n$ trackers downstream, with a factor $\sim0.02^n$.\\
In the case of NA64$\mu$, the level of hermeticity of the detectors is inferred in the plane defined by the muon energy after ECAL and the total energy deposited in the calorimeter (ECAL, VHCAL and HCAL), $(E_\mu';\ E_\text{CAL})$, as shown in Fig. \ref{fig:hermeticity}. Whereas region $A$ corresponds to events with large energy deposit on the HCAL and the diagonal $B$ to events with energy deposited in the ECAL, the bulk $C$ is associated to events with energy deposition in the calorimeters compatible with a MIP. Thus a high level of hermeticity is reached for all events lying within those three regions. On the other side, poor detector hermeticity due to geometrical acceptance or dead material can lead to events with large missing energy, and thus leakage towards the signal box defined in the region $D$ (red box) of Fig. \ref{fig:hermeticity} $(E_\text{CAL}\leq20\ \text{GeV};\ E_\mu'\leq80\ \text{GeV})$. From the distribution extrapolation in the plane $(E_\mu';\ E_\text{CAL})$, the probability of leakage in region $D$ due to non-hermeticity (i.e. detector acceptance) are estimated to be $\lesssim10^{-12}$ per MOT.  \\
Apart from geometrical properties of the detectors, and thus acceptance, two main sources of physical background contribute to events with large missing energy. The first one arises from hadron admixture in the M2 beamline, typically charged and neutral hadrons, such as $\pi^{-}$, $K^{-}$ and $K_{L(s)}^{0}$, and their subsequent (semi-)leptonic decays along the set-up. The level of contamination is measured with a set of Beryllium absorbers in the beamline, and found to be $P_{h} = \pi/\mu\sim 10^{-6}$, with $K/\pi\sim0.03$ \cite{Doble:1994np}. To estimate the hadron decay probability, $P_{h\rightarrow X}$, and the related level of background, hadrons are simulated at the end of the COMPASS BMS to account for particle mis-identification (mis-PID) through momentum reconstruction. From MC simulations, along the typical distance to the active target of $\sim$36 m, it is found that $P_{K^{-}\rightarrow X}\sim\mathcal{O}(10^{-3})$, whereas $P_{\pi^{-},\ K_{L}^{0}\rightarrow X}\sim\mathcal{O}(10^{-4})$. Thus the total number of decay hadrons before the entrance of the set-up is estimated through $N_{h\rightarrow X}=N_\text{MOT}\times P_{h}\times P_{h\rightarrow X}$, which is $\sim\mathcal{O}(10^{-10}-10^{-11})$ per MOT. From those in-flight decays, background is associated to events with final state muons in the decays products, namely $h\rightarrow\mu^{-}X$, where $X$ is mostly associated to neutrino in the case of $\pi^{-}$ and $K^{-}$, susceptible to carry away a large fraction of its parent hadron energy, thus mimicking a $Z'$-strahlung event with missing energy. For such a distance, the probability of kaons decaying to muons through the purely leptonic channel, $K^{-}\rightarrow\mu^{-}\bar{\nu}_\mu$, is about $P_{K^{-}\rightarrow\mu^{-}}\simeq0.018$. For final state muons with energy $E_\mu<100$ GeV, this probability reduces to $P_{K^{-}\rightarrow\mu^{-}}(E_\mu<100\ \text{GeV})\simeq0.011$. In the case of pion decays, this probability is strongly reduced because of the kinematics of the process. The overall probability for such a process, given the kaons contamination of the beam, is estimated at the level of $3.3\times10^{-10}$ per MOT. For the full set of selection criteria, this background reduces to $1.1\times10^{-11}$ per MOT. Such background can be further reduced by the mean of additional absorbers in the M2 beam-line. For a 3.8$\lambda_I$=150 cm aluminium absorber, this probability is reduced by a factor $e^{-3.7}\simeq0.023$. Additionally, using a dedicated magnetic spectrometer just before the active target \cite{Gninenko:2014pea,Banerjee:2019na64mu} reduces the effective hadron decay length to 4 m, thus suppressing further this background by at least an order of magnitude according to simulations.\\
The second important source of background contributing to missing energy events originates from leading hadrons production in the target. Those arise from muon nuclear-interactions, $\mu^{-}N\rightarrow\mu^{-}hX$, within the ECAL material, with the outgoing hadron carrying away a significant fraction of the primary muon energy $(E_{h}\geq80$ GeV). Such events can then leak through (punch-through (PT)) the detector elements downstream the target, with two possible scenarios mimicking a signal event: (i) the low energy outgoing muon is poorly detected and the leading charged hadrons deposits an energy compatible with the one of a MIP ($E_\text{HCAL}\sim2.5$ GeV) in the HCAL module; (ii) the outgoing muon is reconstructed with low energy and the neutral hadron traverses the HCAL modules undetected ($E_\text{HCAL}\sim0.1$ GeV). The probability for leading hadrons production in the target is estimated through MC simulations to be $10^{-6}$ per MOT. Similarly, the probability for punching-through a single or multiples HCAL module(s) is estimated with simulations and compared to available experimental data \cite{DENISOV197362,Aalste1993,PhysRevD.42.759}. In the case of a single module, the effect of punch-through charged/neutral high energy hadrons - mostly $\pi$, $K$ and neutrons $n$ - appears as a peak in the low-energy end of the HCAL energy deposited spectrum. By extrapolating the low-end of the spectrum, the PT probability for a single module corresponds to $\lesssim10^{-2}$ per incoming hadron. A similar analysis extended to two and four modules yields a probability of $\lesssim10^{-6}$ and $\lesssim10^{-11}$ respectively. The overall total probability of producing a leading hadron which subsequently punches through two HCAL modules is thus estimated to be $\lesssim10^{-12}$ per MOT.\\
For completeness to this study, muon electromagnetic interactions within the target also constitutes a possible source of background, especially if visible decays of $Z'$ are inferred (typically $Z'\rightarrow\mu^{+}\mu^{-}$). The main process is dimuons production through the emission of a real photon (Bethe-Heitler mechanism), $\mu^{-}N\rightarrow\mu^{-}N\gamma$; $\gamma\rightarrow\mu^{+}\mu^{-}$. Other mechanisms responsible for such di-leptons production, although more suppressed, are the production of dimuon through a virtual photon (Trident process) or through highly energetic knock-on electrons (see e.g. \cite{Chaudhuri1965,Kelner95,akhiezer1965,Bogdanov:2006kr}). The dimuon yield, suppressed by a factor of $(m_e/m_\mu)^5$ compared to electron bremsstrahlung, is estimated through MC simulations to be $\sim10^{-7}$ per MOT. Because of the phase-space associated to this process, final states muons can be efficiently rejected through the double/triple MIP signature in the HCAL modules, as well as within the tracking detectors. For the typical set of cuts used within this framework, possible background due to dimuons production is rejected at the level of $<10^{-12}$ per MOT.\\
Combining all of the above main sources of background, the experiment is expected to be background-free at the level of $\sim10^{11}$ MOTs (see table \ref{table:table1}).
\begin{table}[h]
  \centering
  \caption{\label{table:table1}Main sources of background and expected background level per muons on target (MOT).}
  \label{parset}
  \begin{tabular*}{\columnwidth}{@{\extracolsep{\fill}}ll@{}}
    \hline
    Source of background & Level per MOT\\
    \hline
    Hadron in-flight decay & $\lesssim10^{-11}$ \\
    Momentum mismatch & $\lesssim10^{-12}$ \\
    Detector non-hermeticity & $\lesssim10^{-12}$\\
    Single-hadron punch-through & $\lesssim10^{-12}$ \\
    Dimuon production & $<10^{-12}$\\ 
    \hline 
    Total (conservatively) & $\lesssim10^{-11}$\\
    \hline
  \end{tabular*}
\end{table}

\section{Future sensitivity to the $(g-2)_\mu$}
\begin{figure}[h]
  \centering
  \includegraphics[width=0.5\textwidth]{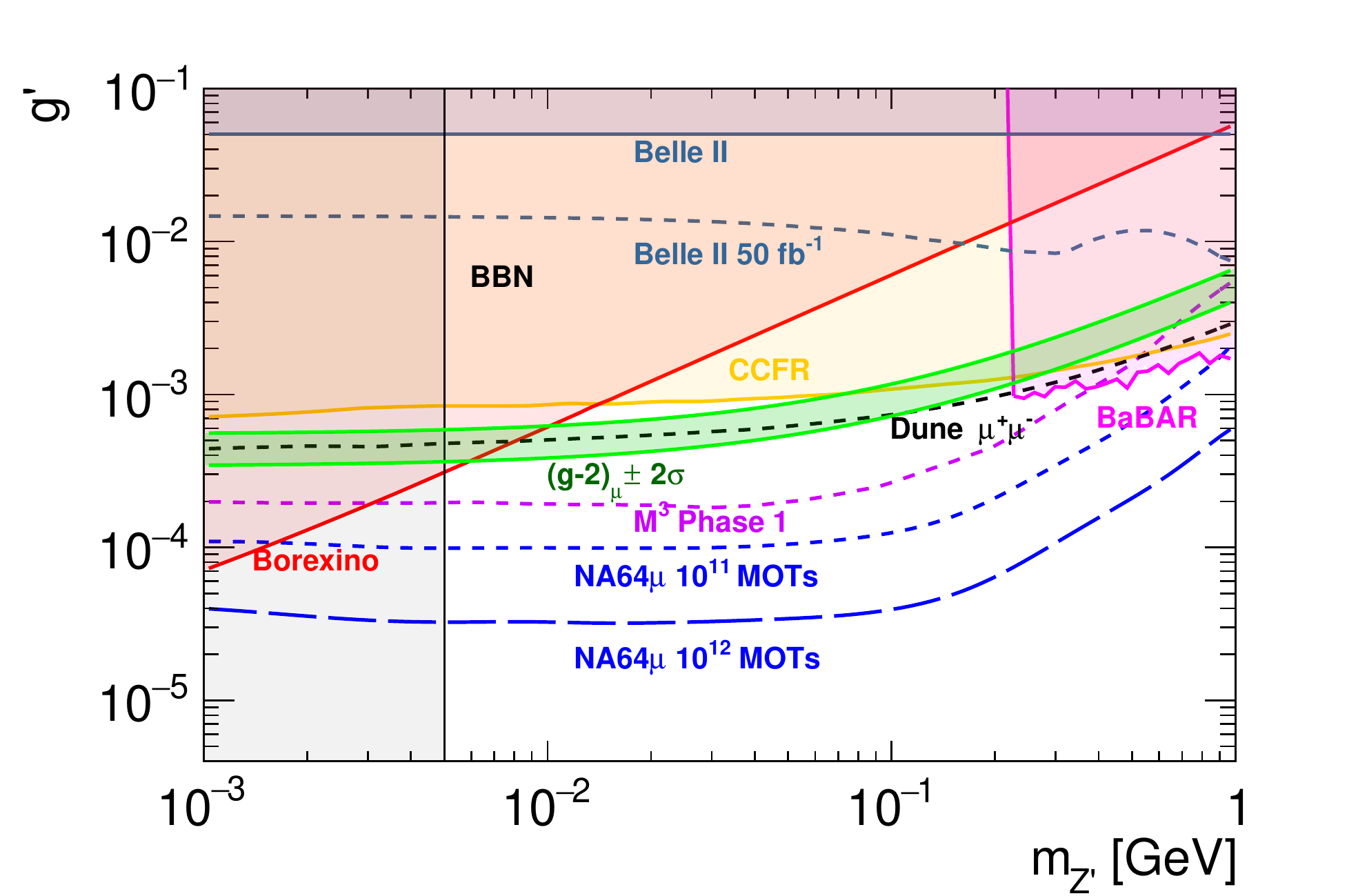}
  \caption{\label{fig:Limits}Projected sensitivities for the invisible mode $Z'\rightarrow\bar{\nu}{\nu}$ in the $(m_{Z'},\ g')$ plane for a total of $10^{11}$ and $10^{12}$ MOT with the NA64$\mu$ experiment. The results are given using the selection criteria (i-iv) and the requirement of the event lying in the signal box. The necessary $(m_{Z'},\ g')$ values to explain the $(g-2)_\mu$ anomaly are shown within the $2\sigma$ green band. Are also shown current experimental constraints from CCFR \cite{PhysRevLett.66.3117,PhysRevLett.113.091801}, BOREXINO \cite{10.1093/ptep/ptx050}, BaBAR \cite{PhysRevD.94.011102} and Belle-II \cite{Belle-II:2019qfb}, together with the cosmological constraints from the Big-Bang Nucleosynthesis (BBN) \cite{PhysRevD.92.113004,PhysRevLett.111.199001,Escudero:2019gzq}. Projected sensitivities from Dune \cite{PhysRevD.100.055012,PhysRevD.100.115029}, Belle-II \cite{BelleII:Campajola2021} and $M^{3}$ \cite{Kahn:2018cqs} are also plotted.}
\end{figure}

The signal yield of invisible decay of $Z'$ to SM neutrinos, $Z'\rightarrow\bar{\nu}\nu$, can be estimated according to \cite{kirpichnikov2021probing,Banerjee:2019na64mu} such that:
\begin{equation}
  \label{eq:signalyield}
  N_{Z'}^{(\bar{\nu}\nu)}=N_\text{MOT}\cdot\frac{\rho\mathcal{N}_A}{A}\cdot\sum_{i}\sigma_\text{tot}(E_\mu^i)\Delta L_i\cdot\text{Br}(Z'\rightarrow\bar{\nu}\nu),
\end{equation}
where $\mathcal{N}_A$ is the Avogadro number, $\rho$ and $A$ correspond to the target material properties, $\Delta L_i$ is the $i$th step length of the muon with energy $E_\mu^{i}$ within the target, and $\sigma_\text{tot}$ the total cross-section for $Z'$ emission. The 90\% C.L. upper limit on $g'$ are calculated using eq. (\ref{eq:signalyield}), thus requiring $N_{Z'}^{(\bar{\nu}\nu)}>2.3$ events, in the $(m_{Z'},\ g')$ plane. The corresponding results are shown Fig. \ref{fig:Limits} for respectively $N_\text{MOT}=10^{11},\ 10^{12}$, assuming the selection criteria (i-iv) and a single scattered muon with energy $E_\mu'\leq80$ GeV. Is also shown the necessary values of $g'$ and $m_{Z'}$ necessary to explain the muon $(g-2)_\mu$ within the $2\sigma$ band. It can be seen that for $N_{MOT}\geq10^{11}$, the parameters space necessary to explain the muon anomalous magnetic moment is fully covered.\\ 
For completeness, existing experimental bounds on $Z'$ from the gauge extension $L_\mu-L_\tau$ theory are shown for neutrino trident production, $\nu N\rightarrow\nu N\mu\mu$, with the CCFR experiment \cite{PhysRevLett.66.3117,PhysRevLett.113.091801}, as well as from neutrino-electron scattering with the BOREXINO experiment \cite{10.1093/ptep/ptx050}. Also shown are experimental constraints from electron-positron colliders with the BaBAR \cite{PhysRevD.94.011102} and Belle-II \cite{Belle-II:2019qfb} experiments. As comparison, projected sensitivities for Dune \cite{PhysRevD.100.055012,PhysRevD.100.115029}, Belle-II \cite{BelleII:Campajola2021} and $M^{3}$ \cite{Kahn:2018cqs} are plotted alongside our estimated limits. 

\section{Conclusion}
In this work, we presented the expected sensitivity of NA64$\mu$ experiment to look for a light $Z'$ boson coupled to muons as a remaining low mass explanation of the $(g-2)_\mu$ muon anomaly. The minimal model based on the broken $U(1)_{L_{\mu}-L_{\tau}}$ symmetry is used as benchmark in these studies. We focused on the optimisation and design of the experimental setup for the first phase of the experiment dedicated to demonstrate the feasibility of the technique. The trigger efficiency and the detector hermeticity has been studied using a dedicated \texttt{GEANT4}-based MC simulation framework and a realistic M2 beam-optics simulation. A trigger using the scattered muon deflection after traversing the magnet spectrometer has been designed to keep the primary beam below $0.1\%$ and a signal efficiency between $15-35\%$ mass dependent. The low trigger efficiency for smaller masses is compensated by the mass dependence of the cross-section. The main expected background sources arise from momentum mis-reconstruction in the two magnet spectrometers, hadron contamination in the beam line and detector hermeticity. The results obtained from the simulation show that the background level is below $10^{-11}$ per MOT being dominated by the decay in flight of the remaining hadron contamination in the M2 beamline. We also showed two methods which can potentially reduce this background by at least an order of magnitude. Finally, we studied the experiment projected sensitivities compared to present and future experiments aiming to explore similar searches. These first estimates based on simulations reveal that with 10$^{11}$ MOT we can probe the region relevant to the $(g-2)_\mu$ anomaly, obtaining the most sensitive coverage for masses below 200 MeV. The presented simulation results and calculations are to be validated in the scheduled pilot run.
%If the scheduled pilot run will validate the simulation results, NA64$\mu$ will be able to explore all the remaining phase space which could provide an explanation for the $g-2$ muon anomaly.

\section*{Aknowledgements}
We acknowledge the members of the NA64 collaboration for fruitful discussions, 
in particular, N. V. Krasnikov. The work of P. Crivelli, E. Depero, L. Molina Bueno and H. Sieber is supported by ETH Zürich and 
SNSF Grant No. 169133, 186181, 186158 and 197346 (Switzerland). The work of D. V. Kirpichnikov on MC simulation of $Z'$ emission is supported by the Russian Science Foundation RSF grant 21-12-00379.

%\clearpage
\bibliographystyle{apsrev4-2}
\bibliography{../Bibliography/bibliographyNA64_inspiresFormat.bib,../Bibliography/bibliographyNA64exp_inspiresFormat.bib,../Bibliography/bibliographyOther_inspiresFormat.bib}

\begin{thebibliography}{71}%
\makeatletter
\providecommand \@ifxundefined [1]{%
 \@ifx{#1\undefined}
}%
\providecommand \@ifnum [1]{%
 \ifnum #1\expandafter \@firstoftwo
 \else \expandafter \@secondoftwo
 \fi
}%
\providecommand \@ifx [1]{%
 \ifx #1\expandafter \@firstoftwo
 \else \expandafter \@secondoftwo
 \fi
}%
\providecommand \natexlab [1]{#1}%
\providecommand \enquote  [1]{``#1''}%
\providecommand \bibnamefont  [1]{#1}%
\providecommand \bibfnamefont [1]{#1}%
\providecommand \citenamefont [1]{#1}%
\providecommand \href@noop [0]{\@secondoftwo}%
\providecommand \href [0]{\begingroup \@sanitize@url \@href}%
\providecommand \@href[1]{\@@startlink{#1}\@@href}%
\providecommand \@@href[1]{\endgroup#1\@@endlink}%
\providecommand \@sanitize@url [0]{\catcode `\\12\catcode `\$12\catcode
  `\&12\catcode `\#12\catcode `\^12\catcode `\_12\catcode `\%12\relax}%
\providecommand \@@startlink[1]{}%
\providecommand \@@endlink[0]{}%
\providecommand \url  [0]{\begingroup\@sanitize@url \@url }%
\providecommand \@url [1]{\endgroup\@href {#1}{\urlprefix }}%
\providecommand \urlprefix  [0]{URL }%
\providecommand \Eprint [0]{\href }%
\providecommand \doibase [0]{https://doi.org/}%
\providecommand \selectlanguage [0]{\@gobble}%
\providecommand \bibinfo  [0]{\@secondoftwo}%
\providecommand \bibfield  [0]{\@secondoftwo}%
\providecommand \translation [1]{[#1]}%
\providecommand \BibitemOpen [0]{}%
\providecommand \bibitemStop [0]{}%
\providecommand \bibitemNoStop [0]{.\EOS\space}%
\providecommand \EOS [0]{\spacefactor3000\relax}%
\providecommand \BibitemShut  [1]{\csname bibitem#1\endcsname}%
\let\auto@bib@innerbib\@empty
%</preamble>
\bibitem [{\citenamefont {Abi}\ \emph {et~al.}(2021)\citenamefont {Abi} \emph
  {et~al.}}]{Abi:2021gix}%
  \BibitemOpen
  \bibfield  {author} {\bibinfo {author} {\bibfnamefont {B.}~\bibnamefont
  {Abi}} \emph {et~al.} (\bibinfo {collaboration} {Muon g-2}),\ }\href
  {https://doi.org/10.1103/PhysRevLett.126.141801} {\bibfield  {journal}
  {\bibinfo  {journal} {Phys. Rev. Lett.}\ }\textbf {\bibinfo {volume} {126}},\
  \bibinfo {pages} {141801} (\bibinfo {year} {2021})},\ \Eprint
  {https://arxiv.org/abs/2104.03281} {arXiv:2104.03281 [hep-ex]} \BibitemShut
  {NoStop}%
\bibitem [{\citenamefont {Aoyama}\ \emph {et~al.}(2020)\citenamefont {Aoyama}
  \emph {et~al.}}]{Aoyama:2020ynm}%
  \BibitemOpen
  \bibfield  {author} {\bibinfo {author} {\bibfnamefont {T.}~\bibnamefont
  {Aoyama}} \emph {et~al.},\ }\href
  {https://doi.org/10.1016/j.physrep.2020.07.006} {\bibfield  {journal}
  {\bibinfo  {journal} {Phys. Rept.}\ }\textbf {\bibinfo {volume} {887}},\
  \bibinfo {pages} {1} (\bibinfo {year} {2020})},\ \Eprint
  {https://arxiv.org/abs/2006.04822} {arXiv:2006.04822 [hep-ph]} \BibitemShut
  {NoStop}%
\bibitem [{\citenamefont {Aoyama}\ \emph {et~al.}(2012)\citenamefont {Aoyama},
  \citenamefont {Hayakawa}, \citenamefont {Kinoshita},\ and\ \citenamefont
  {Nio}}]{Aoyama:2012wk}%
  \BibitemOpen
  \bibfield  {author} {\bibinfo {author} {\bibfnamefont {T.}~\bibnamefont
  {Aoyama}}, \bibinfo {author} {\bibfnamefont {M.}~\bibnamefont {Hayakawa}},
  \bibinfo {author} {\bibfnamefont {T.}~\bibnamefont {Kinoshita}},\ and\
  \bibinfo {author} {\bibfnamefont {M.}~\bibnamefont {Nio}},\ }\href
  {https://doi.org/10.1103/PhysRevLett.109.111808} {\bibfield  {journal}
  {\bibinfo  {journal} {Phys. Rev. Lett.}\ }\textbf {\bibinfo {volume} {109}},\
  \bibinfo {pages} {111808} (\bibinfo {year} {2012})},\ \Eprint
  {https://arxiv.org/abs/1205.5370} {arXiv:1205.5370 [hep-ph]} \BibitemShut
  {NoStop}%
\bibitem [{\citenamefont {Aoyama}\ \emph {et~al.}(2019)\citenamefont {Aoyama},
  \citenamefont {Kinoshita},\ and\ \citenamefont {Nio}}]{Aoyama:2019ryr}%
  \BibitemOpen
  \bibfield  {author} {\bibinfo {author} {\bibfnamefont {T.}~\bibnamefont
  {Aoyama}}, \bibinfo {author} {\bibfnamefont {T.}~\bibnamefont {Kinoshita}},\
  and\ \bibinfo {author} {\bibfnamefont {M.}~\bibnamefont {Nio}},\ }\href
  {https://doi.org/10.3390/atoms7010028} {\bibfield  {journal} {\bibinfo
  {journal} {Atoms}\ }\textbf {\bibinfo {volume} {7}},\ \bibinfo {pages} {28}
  (\bibinfo {year} {2019})}\BibitemShut {NoStop}%
\bibitem [{\citenamefont {Czarnecki}\ \emph {et~al.}(2003)\citenamefont
  {Czarnecki}, \citenamefont {Marciano},\ and\ \citenamefont
  {Vainshtein}}]{Czarnecki:2002nt}%
  \BibitemOpen
  \bibfield  {author} {\bibinfo {author} {\bibfnamefont {A.}~\bibnamefont
  {Czarnecki}}, \bibinfo {author} {\bibfnamefont {W.~J.}\ \bibnamefont
  {Marciano}},\ and\ \bibinfo {author} {\bibfnamefont {A.}~\bibnamefont
  {Vainshtein}},\ }\href {https://doi.org/10.1103/PhysRevD.67.073006}
  {\bibfield  {journal} {\bibinfo  {journal} {Phys. Rev.}\ }\textbf {\bibinfo
  {volume} {D67}},\ \bibinfo {pages} {073006} (\bibinfo {year} {2003})},\
  \bibinfo {note} {[Erratum: Phys. Rev. {\bf D73}, 119901 (2006)]},\ \Eprint
  {https://arxiv.org/abs/hep-ph/0212229} {arXiv:hep-ph/0212229 [hep-ph]}
  \BibitemShut {NoStop}%
\bibitem [{\citenamefont {Gnendiger}\ \emph {et~al.}(2013)\citenamefont
  {Gnendiger}, \citenamefont {St{\"o}ckinger},\ and\ \citenamefont
  {St{\"o}ckinger-Kim}}]{Gnendiger:2013pva}%
  \BibitemOpen
  \bibfield  {author} {\bibinfo {author} {\bibfnamefont {C.}~\bibnamefont
  {Gnendiger}}, \bibinfo {author} {\bibfnamefont {D.}~\bibnamefont
  {St{\"o}ckinger}},\ and\ \bibinfo {author} {\bibfnamefont {H.}~\bibnamefont
  {St{\"o}ckinger-Kim}},\ }\href {https://doi.org/10.1103/PhysRevD.88.053005}
  {\bibfield  {journal} {\bibinfo  {journal} {Phys. Rev.}\ }\textbf {\bibinfo
  {volume} {D88}},\ \bibinfo {pages} {053005} (\bibinfo {year} {2013})},\
  \Eprint {https://arxiv.org/abs/1306.5546} {arXiv:1306.5546 [hep-ph]}
  \BibitemShut {NoStop}%
\bibitem [{\citenamefont {Davier}\ \emph {et~al.}(2017)\citenamefont {Davier},
  \citenamefont {Hoecker}, \citenamefont {Malaescu},\ and\ \citenamefont
  {Zhang}}]{Davier:2017zfy}%
  \BibitemOpen
  \bibfield  {author} {\bibinfo {author} {\bibfnamefont {M.}~\bibnamefont
  {Davier}}, \bibinfo {author} {\bibfnamefont {A.}~\bibnamefont {Hoecker}},
  \bibinfo {author} {\bibfnamefont {B.}~\bibnamefont {Malaescu}},\ and\
  \bibinfo {author} {\bibfnamefont {Z.}~\bibnamefont {Zhang}},\ }\href
  {https://doi.org/10.1140/epjc/s10052-017-5161-6} {\bibfield  {journal}
  {\bibinfo  {journal} {Eur. Phys. J.}\ }\textbf {\bibinfo {volume} {C77}},\
  \bibinfo {pages} {827} (\bibinfo {year} {2017})},\ \Eprint
  {https://arxiv.org/abs/1706.09436} {arXiv:1706.09436 [hep-ph]} \BibitemShut
  {NoStop}%
\bibitem [{\citenamefont {Keshavarzi}\ \emph {et~al.}(2018)\citenamefont
  {Keshavarzi}, \citenamefont {Nomura},\ and\ \citenamefont
  {Teubner}}]{Keshavarzi:2018mgv}%
  \BibitemOpen
  \bibfield  {author} {\bibinfo {author} {\bibfnamefont {A.}~\bibnamefont
  {Keshavarzi}}, \bibinfo {author} {\bibfnamefont {D.}~\bibnamefont {Nomura}},\
  and\ \bibinfo {author} {\bibfnamefont {T.}~\bibnamefont {Teubner}},\ }\href
  {https://doi.org/10.1103/PhysRevD.97.114025} {\bibfield  {journal} {\bibinfo
  {journal} {Phys. Rev.}\ }\textbf {\bibinfo {volume} {D97}},\ \bibinfo {pages}
  {114025} (\bibinfo {year} {2018})},\ \Eprint
  {https://arxiv.org/abs/1802.02995} {arXiv:1802.02995 [hep-ph]} \BibitemShut
  {NoStop}%
\bibitem [{\citenamefont {Colangelo}\ \emph {et~al.}(2019)\citenamefont
  {Colangelo}, \citenamefont {Hoferichter},\ and\ \citenamefont
  {Stoffer}}]{Colangelo:2018mtw}%
  \BibitemOpen
  \bibfield  {author} {\bibinfo {author} {\bibfnamefont {G.}~\bibnamefont
  {Colangelo}}, \bibinfo {author} {\bibfnamefont {M.}~\bibnamefont
  {Hoferichter}},\ and\ \bibinfo {author} {\bibfnamefont {P.}~\bibnamefont
  {Stoffer}},\ }\href {https://doi.org/10.1007/JHEP02(2019)006} {\bibfield
  {journal} {\bibinfo  {journal} {JHEP}\ }\textbf {\bibinfo {volume} {02}},\
  \bibinfo {pages} {006}},\ \Eprint {https://arxiv.org/abs/1810.00007}
  {arXiv:1810.00007 [hep-ph]} \BibitemShut {NoStop}%
\bibitem [{\citenamefont {Hoferichter}\ \emph {et~al.}(2019)\citenamefont
  {Hoferichter}, \citenamefont {Hoid},\ and\ \citenamefont
  {Kubis}}]{Hoferichter:2019gzf}%
  \BibitemOpen
  \bibfield  {author} {\bibinfo {author} {\bibfnamefont {M.}~\bibnamefont
  {Hoferichter}}, \bibinfo {author} {\bibfnamefont {B.-L.}\ \bibnamefont
  {Hoid}},\ and\ \bibinfo {author} {\bibfnamefont {B.}~\bibnamefont {Kubis}},\
  }\href {https://doi.org/10.1007/JHEP08(2019)137} {\bibfield  {journal}
  {\bibinfo  {journal} {JHEP}\ }\textbf {\bibinfo {volume} {08}},\ \bibinfo
  {pages} {137}},\ \Eprint {https://arxiv.org/abs/1907.01556} {arXiv:1907.01556
  [hep-ph]} \BibitemShut {NoStop}%
\bibitem [{\citenamefont {Davier}\ \emph {et~al.}(2020)\citenamefont {Davier},
  \citenamefont {Hoecker}, \citenamefont {Malaescu},\ and\ \citenamefont
  {Zhang}}]{Davier:2019can}%
  \BibitemOpen
  \bibfield  {author} {\bibinfo {author} {\bibfnamefont {M.}~\bibnamefont
  {Davier}}, \bibinfo {author} {\bibfnamefont {A.}~\bibnamefont {Hoecker}},
  \bibinfo {author} {\bibfnamefont {B.}~\bibnamefont {Malaescu}},\ and\
  \bibinfo {author} {\bibfnamefont {Z.}~\bibnamefont {Zhang}},\ }\href
  {https://doi.org/10.1140/epjc/s10052-020-7792-2} {\bibfield  {journal}
  {\bibinfo  {journal} {Eur. Phys. J.}\ }\textbf {\bibinfo {volume} {C80}},\
  \bibinfo {pages} {241} (\bibinfo {year} {2020})},\ \bibinfo {note} {[Erratum:
  Eur. Phys. J. {\bf C80}, 410 (2020)]},\ \Eprint
  {https://arxiv.org/abs/1908.00921} {arXiv:1908.00921 [hep-ph]} \BibitemShut
  {NoStop}%
\bibitem [{\citenamefont {Keshavarzi}\ \emph {et~al.}(2020)\citenamefont
  {Keshavarzi}, \citenamefont {Nomura},\ and\ \citenamefont
  {Teubner}}]{Keshavarzi:2019abf}%
  \BibitemOpen
  \bibfield  {author} {\bibinfo {author} {\bibfnamefont {A.}~\bibnamefont
  {Keshavarzi}}, \bibinfo {author} {\bibfnamefont {D.}~\bibnamefont {Nomura}},\
  and\ \bibinfo {author} {\bibfnamefont {T.}~\bibnamefont {Teubner}},\ }\href
  {https://doi.org/10.1103/PhysRevD.101.014029} {\bibfield  {journal} {\bibinfo
   {journal} {Phys. Rev.}\ }\textbf {\bibinfo {volume} {D101}},\ \bibinfo
  {pages} {014029} (\bibinfo {year} {2020})},\ \Eprint
  {https://arxiv.org/abs/1911.00367} {arXiv:1911.00367 [hep-ph]} \BibitemShut
  {NoStop}%
\bibitem [{\citenamefont {Kurz}\ \emph {et~al.}(2014)\citenamefont {Kurz},
  \citenamefont {Liu}, \citenamefont {Marquard},\ and\ \citenamefont
  {Steinhauser}}]{Kurz:2014wya}%
  \BibitemOpen
  \bibfield  {author} {\bibinfo {author} {\bibfnamefont {A.}~\bibnamefont
  {Kurz}}, \bibinfo {author} {\bibfnamefont {T.}~\bibnamefont {Liu}}, \bibinfo
  {author} {\bibfnamefont {P.}~\bibnamefont {Marquard}},\ and\ \bibinfo
  {author} {\bibfnamefont {M.}~\bibnamefont {Steinhauser}},\ }\href
  {https://doi.org/10.1016/j.physletb.2014.05.043} {\bibfield  {journal}
  {\bibinfo  {journal} {Phys. Lett.}\ }\textbf {\bibinfo {volume} {B734}},\
  \bibinfo {pages} {144} (\bibinfo {year} {2014})},\ \Eprint
  {https://arxiv.org/abs/1403.6400} {arXiv:1403.6400 [hep-ph]} \BibitemShut
  {NoStop}%
\bibitem [{\citenamefont {Melnikov}\ and\ \citenamefont
  {Vainshtein}(2004)}]{Melnikov:2003xd}%
  \BibitemOpen
  \bibfield  {author} {\bibinfo {author} {\bibfnamefont {K.}~\bibnamefont
  {Melnikov}}\ and\ \bibinfo {author} {\bibfnamefont {A.}~\bibnamefont
  {Vainshtein}},\ }\href {https://doi.org/10.1103/PhysRevD.70.113006}
  {\bibfield  {journal} {\bibinfo  {journal} {Phys. Rev.}\ }\textbf {\bibinfo
  {volume} {D70}},\ \bibinfo {pages} {113006} (\bibinfo {year} {2004})},\
  \Eprint {https://arxiv.org/abs/hep-ph/0312226} {arXiv:hep-ph/0312226
  [hep-ph]} \BibitemShut {NoStop}%
\bibitem [{\citenamefont {Masjuan}\ and\ \citenamefont
  {S{\'a}nchez-Puertas}(2017)}]{Masjuan:2017tvw}%
  \BibitemOpen
  \bibfield  {author} {\bibinfo {author} {\bibfnamefont {P.}~\bibnamefont
  {Masjuan}}\ and\ \bibinfo {author} {\bibfnamefont {P.}~\bibnamefont
  {S{\'a}nchez-Puertas}},\ }\href {https://doi.org/10.1103/PhysRevD.95.054026}
  {\bibfield  {journal} {\bibinfo  {journal} {Phys. Rev.}\ }\textbf {\bibinfo
  {volume} {D95}},\ \bibinfo {pages} {054026} (\bibinfo {year} {2017})},\
  \Eprint {https://arxiv.org/abs/1701.05829} {arXiv:1701.05829 [hep-ph]}
  \BibitemShut {NoStop}%
\bibitem [{\citenamefont {Colangelo}\ \emph {et~al.}(2017)\citenamefont
  {Colangelo}, \citenamefont {Hoferichter}, \citenamefont {Procura},\ and\
  \citenamefont {Stoffer}}]{Colangelo:2017fiz}%
  \BibitemOpen
  \bibfield  {author} {\bibinfo {author} {\bibfnamefont {G.}~\bibnamefont
  {Colangelo}}, \bibinfo {author} {\bibfnamefont {M.}~\bibnamefont
  {Hoferichter}}, \bibinfo {author} {\bibfnamefont {M.}~\bibnamefont
  {Procura}},\ and\ \bibinfo {author} {\bibfnamefont {P.}~\bibnamefont
  {Stoffer}},\ }\href {https://doi.org/10.1007/JHEP04(2017)161} {\bibfield
  {journal} {\bibinfo  {journal} {JHEP}\ }\textbf {\bibinfo {volume} {04}},\
  \bibinfo {pages} {161}},\ \Eprint {https://arxiv.org/abs/1702.07347}
  {arXiv:1702.07347 [hep-ph]} \BibitemShut {NoStop}%
\bibitem [{\citenamefont {Hoferichter}\ \emph {et~al.}(2018)\citenamefont
  {Hoferichter}, \citenamefont {Hoid}, \citenamefont {Kubis}, \citenamefont
  {Leupold},\ and\ \citenamefont {Schneider}}]{Hoferichter:2018kwz}%
  \BibitemOpen
  \bibfield  {author} {\bibinfo {author} {\bibfnamefont {M.}~\bibnamefont
  {Hoferichter}}, \bibinfo {author} {\bibfnamefont {B.-L.}\ \bibnamefont
  {Hoid}}, \bibinfo {author} {\bibfnamefont {B.}~\bibnamefont {Kubis}},
  \bibinfo {author} {\bibfnamefont {S.}~\bibnamefont {Leupold}},\ and\ \bibinfo
  {author} {\bibfnamefont {S.~P.}\ \bibnamefont {Schneider}},\ }\href
  {https://doi.org/10.1007/JHEP10(2018)141} {\bibfield  {journal} {\bibinfo
  {journal} {JHEP}\ }\textbf {\bibinfo {volume} {10}},\ \bibinfo {pages}
  {141}},\ \Eprint {https://arxiv.org/abs/1808.04823} {arXiv:1808.04823
  [hep-ph]} \BibitemShut {NoStop}%
\bibitem [{\citenamefont {G{\'e}rardin}\ \emph {et~al.}(2019)\citenamefont
  {G{\'e}rardin}, \citenamefont {Meyer},\ and\ \citenamefont
  {Nyffeler}}]{Gerardin:2019vio}%
  \BibitemOpen
  \bibfield  {author} {\bibinfo {author} {\bibfnamefont {A.}~\bibnamefont
  {G{\'e}rardin}}, \bibinfo {author} {\bibfnamefont {H.~B.}\ \bibnamefont
  {Meyer}},\ and\ \bibinfo {author} {\bibfnamefont {A.}~\bibnamefont
  {Nyffeler}},\ }\href {https://doi.org/10.1103/PhysRevD.100.034520} {\bibfield
   {journal} {\bibinfo  {journal} {Phys. Rev.}\ }\textbf {\bibinfo {volume}
  {D100}},\ \bibinfo {pages} {034520} (\bibinfo {year} {2019})},\ \Eprint
  {https://arxiv.org/abs/1903.09471} {arXiv:1903.09471 [hep-lat]} \BibitemShut
  {NoStop}%
\bibitem [{\citenamefont {Bijnens}\ \emph {et~al.}(2019)\citenamefont
  {Bijnens}, \citenamefont {Hermansson-Truedsson},\ and\ \citenamefont
  {Rodr{\'i}guez-S{\'a}nchez}}]{Bijnens:2019ghy}%
  \BibitemOpen
  \bibfield  {author} {\bibinfo {author} {\bibfnamefont {J.}~\bibnamefont
  {Bijnens}}, \bibinfo {author} {\bibfnamefont {N.}~\bibnamefont
  {Hermansson-Truedsson}},\ and\ \bibinfo {author} {\bibfnamefont
  {A.}~\bibnamefont {Rodr{\'i}guez-S{\'a}nchez}},\ }\href
  {https://doi.org/10.1016/j.physletb.2019.134994} {\bibfield  {journal}
  {\bibinfo  {journal} {Phys. Lett.}\ }\textbf {\bibinfo {volume} {B798}},\
  \bibinfo {pages} {134994} (\bibinfo {year} {2019})},\ \Eprint
  {https://arxiv.org/abs/1908.03331} {arXiv:1908.03331 [hep-ph]} \BibitemShut
  {NoStop}%
\bibitem [{\citenamefont {Colangelo}\ \emph {et~al.}(2020)\citenamefont
  {Colangelo}, \citenamefont {Hagelstein}, \citenamefont {Hoferichter},
  \citenamefont {Laub},\ and\ \citenamefont {Stoffer}}]{Colangelo:2019uex}%
  \BibitemOpen
  \bibfield  {author} {\bibinfo {author} {\bibfnamefont {G.}~\bibnamefont
  {Colangelo}}, \bibinfo {author} {\bibfnamefont {F.}~\bibnamefont
  {Hagelstein}}, \bibinfo {author} {\bibfnamefont {M.}~\bibnamefont
  {Hoferichter}}, \bibinfo {author} {\bibfnamefont {L.}~\bibnamefont {Laub}},\
  and\ \bibinfo {author} {\bibfnamefont {P.}~\bibnamefont {Stoffer}},\ }\href
  {https://doi.org/10.1007/JHEP03(2020)101} {\bibfield  {journal} {\bibinfo
  {journal} {JHEP}\ }\textbf {\bibinfo {volume} {03}},\ \bibinfo {pages}
  {101}},\ \Eprint {https://arxiv.org/abs/1910.13432} {arXiv:1910.13432
  [hep-ph]} \BibitemShut {NoStop}%
\bibitem [{\citenamefont {Blum}\ \emph {et~al.}(2020)\citenamefont {Blum},
  \citenamefont {Christ}, \citenamefont {Hayakawa}, \citenamefont {Izubuchi},
  \citenamefont {Jin}, \citenamefont {Jung},\ and\ \citenamefont
  {Lehner}}]{Blum:2019ugy}%
  \BibitemOpen
  \bibfield  {author} {\bibinfo {author} {\bibfnamefont {T.}~\bibnamefont
  {Blum}}, \bibinfo {author} {\bibfnamefont {N.}~\bibnamefont {Christ}},
  \bibinfo {author} {\bibfnamefont {M.}~\bibnamefont {Hayakawa}}, \bibinfo
  {author} {\bibfnamefont {T.}~\bibnamefont {Izubuchi}}, \bibinfo {author}
  {\bibfnamefont {L.}~\bibnamefont {Jin}}, \bibinfo {author} {\bibfnamefont
  {C.}~\bibnamefont {Jung}},\ and\ \bibinfo {author} {\bibfnamefont
  {C.}~\bibnamefont {Lehner}},\ }\href
  {https://doi.org/10.1103/PhysRevLett.124.132002} {\bibfield  {journal}
  {\bibinfo  {journal} {Phys. Rev. Lett.}\ }\textbf {\bibinfo {volume} {124}},\
  \bibinfo {pages} {132002} (\bibinfo {year} {2020})},\ \Eprint
  {https://arxiv.org/abs/1911.08123} {arXiv:1911.08123 [hep-lat]} \BibitemShut
  {NoStop}%
\bibitem [{\citenamefont {Colangelo}\ \emph {et~al.}(2014)\citenamefont
  {Colangelo}, \citenamefont {Hoferichter}, \citenamefont {Nyffeler},
  \citenamefont {Passera},\ and\ \citenamefont {Stoffer}}]{Colangelo:2014qya}%
  \BibitemOpen
  \bibfield  {author} {\bibinfo {author} {\bibfnamefont {G.}~\bibnamefont
  {Colangelo}}, \bibinfo {author} {\bibfnamefont {M.}~\bibnamefont
  {Hoferichter}}, \bibinfo {author} {\bibfnamefont {A.}~\bibnamefont
  {Nyffeler}}, \bibinfo {author} {\bibfnamefont {M.}~\bibnamefont {Passera}},\
  and\ \bibinfo {author} {\bibfnamefont {P.}~\bibnamefont {Stoffer}},\ }\href
  {https://doi.org/10.1016/j.physletb.2014.06.012} {\bibfield  {journal}
  {\bibinfo  {journal} {Phys. Lett.}\ }\textbf {\bibinfo {volume} {B735}},\
  \bibinfo {pages} {90} (\bibinfo {year} {2014})},\ \Eprint
  {https://arxiv.org/abs/1403.7512} {arXiv:1403.7512 [hep-ph]} \BibitemShut
  {NoStop}%
%%CITATION = ARXIV:1403.7512;%%
\bibitem [{\citenamefont {Lindner}\ \emph {et~al.}(2018)\citenamefont {Lindner}
  \emph {et~al.}}]{LINDNER20181}%
  \BibitemOpen
  \bibfield  {author} {\bibinfo {author} {\bibfnamefont {M.}~\bibnamefont
  {Lindner}} \emph {et~al.},\ }\href
  {https://doi.org/https://doi.org/10.1016/j.physrep.2017.12.001} {\bibfield
  {journal} {\bibinfo  {journal} {Phys. Rept.}\ }\textbf {\bibinfo {volume}
  {731}},\ \bibinfo {pages} {1} (\bibinfo {year} {2018})}\BibitemShut {NoStop}%
\bibitem [{\citenamefont {St\"ockinger}(2013)}]{Stockinger:2013rna}%
  \BibitemOpen
  \bibfield  {author} {\bibinfo {author} {\bibfnamefont {D.}~\bibnamefont
  {St\"ockinger}},\ }\href {https://doi.org/10.1007/s10751-013-0804-y}
  {\bibfield  {journal} {\bibinfo  {journal} {Hyperfine Interact.}\ }\textbf
  {\bibinfo {volume} {214}},\ \bibinfo {pages} {13} (\bibinfo {year}
  {2013})}\BibitemShut {NoStop}%
\bibitem [{\citenamefont {Miller}\ \emph {et~al.}(2012)\citenamefont {Miller},
  \citenamefont {Eduardo~de}, \citenamefont {Roberts},\ and\ \citenamefont
  {St\"ackinger}}]{doi:10.1146/annurev-nucl-031312-120340}%
  \BibitemOpen
  \bibfield  {author} {\bibinfo {author} {\bibfnamefont {J.~P.}\ \bibnamefont
  {Miller}}, \bibinfo {author} {\bibfnamefont {R.}~\bibnamefont {Eduardo~de}},
  \bibinfo {author} {\bibfnamefont {B.~L.}\ \bibnamefont {Roberts}},\ and\
  \bibinfo {author} {\bibfnamefont {D.}~\bibnamefont {St\"ackinger}},\ }\href
  {https://doi.org/10.1146/annurev-nucl-031312-120340} {\bibfield  {journal}
  {\bibinfo  {journal} {Ann. Rev. Nucl. Part. Sci.}\ }\textbf {\bibinfo
  {volume} {62}},\ \bibinfo {pages} {237} (\bibinfo {year} {2012})},\ \Eprint
  {https://arxiv.org/abs/https://doi.org/10.1146/annurev-nucl-031312-120340}
  {https://doi.org/10.1146/annurev-nucl-031312-120340} \BibitemShut {NoStop}%
\bibitem [{\citenamefont {Gninenko}\ and\ \citenamefont
  {Krasnikov}(2001)}]{Gninenko:2001hx}%
  \BibitemOpen
  \bibfield  {author} {\bibinfo {author} {\bibfnamefont {S.~N.}\ \bibnamefont
  {Gninenko}}\ and\ \bibinfo {author} {\bibfnamefont {N.~V.}\ \bibnamefont
  {Krasnikov}},\ }\href {https://doi.org/10.1016/S0370-2693(01)00693-1}
  {\bibfield  {journal} {\bibinfo  {journal} {Phys. Lett. B}\ }\textbf
  {\bibinfo {volume} {513}},\ \bibinfo {pages} {119} (\bibinfo {year}
  {2001})},\ \Eprint {https://arxiv.org/abs/hep-ph/0102222}
  {arXiv:hep-ph/0102222} \BibitemShut {NoStop}%
\bibitem [{\citenamefont {Gninenko}\ \emph {et~al.}(2015)\citenamefont
  {Gninenko}, \citenamefont {Krasnikov},\ and\ \citenamefont
  {Matveev}}]{Gninenko:2014pea}%
  \BibitemOpen
  \bibfield  {author} {\bibinfo {author} {\bibfnamefont {S.~N.}\ \bibnamefont
  {Gninenko}}, \bibinfo {author} {\bibfnamefont {N.~V.}\ \bibnamefont
  {Krasnikov}},\ and\ \bibinfo {author} {\bibfnamefont {V.~A.}\ \bibnamefont
  {Matveev}},\ }\href {https://doi.org/10.1103/PhysRevD.91.095015} {\bibfield
  {journal} {\bibinfo  {journal} {Phys. Rev. D}\ }\textbf {\bibinfo {volume}
  {91}},\ \bibinfo {pages} {095015} (\bibinfo {year} {2015})},\ \Eprint
  {https://arxiv.org/abs/1412.1400} {arXiv:1412.1400 [hep-ph]} \BibitemShut
  {NoStop}%
\bibitem [{\citenamefont {Chen}\ \emph {et~al.}(2017)\citenamefont {Chen},
  \citenamefont {Pospelov},\ and\ \citenamefont {Zhong}}]{Chen:2017awl}%
  \BibitemOpen
  \bibfield  {author} {\bibinfo {author} {\bibfnamefont {C.-Y.}\ \bibnamefont
  {Chen}}, \bibinfo {author} {\bibfnamefont {M.}~\bibnamefont {Pospelov}},\
  and\ \bibinfo {author} {\bibfnamefont {Y.-M.}\ \bibnamefont {Zhong}},\ }\href
  {https://doi.org/10.1103/PhysRevD.95.115005} {\bibfield  {journal} {\bibinfo
  {journal} {Phys. Rev. D}\ }\textbf {\bibinfo {volume} {95}},\ \bibinfo
  {pages} {115005} (\bibinfo {year} {2017})},\ \Eprint
  {https://arxiv.org/abs/1701.07437} {arXiv:1701.07437 [hep-ph]} \BibitemShut
  {NoStop}%
\bibitem [{\citenamefont {Gninenko}\ and\ \citenamefont
  {Krasnikov}(2018)}]{Gninenko:2018tlp}%
  \BibitemOpen
  \bibfield  {author} {\bibinfo {author} {\bibfnamefont {S.~N.}\ \bibnamefont
  {Gninenko}}\ and\ \bibinfo {author} {\bibfnamefont {N.~V.}\ \bibnamefont
  {Krasnikov}},\ }\href {https://doi.org/10.1016/j.physletb.2018.06.043}
  {\bibfield  {journal} {\bibinfo  {journal} {Phys. Lett. B}\ }\textbf
  {\bibinfo {volume} {783}},\ \bibinfo {pages} {24} (\bibinfo {year} {2018})},\
  \Eprint {https://arxiv.org/abs/1801.10448} {arXiv:1801.10448 [hep-ph]}
  \BibitemShut {NoStop}%
\bibitem [{\citenamefont {Kirpichnikov}\ \emph {et~al.}(2020)\citenamefont
  {Kirpichnikov}, \citenamefont {Lyubovitskij},\ and\ \citenamefont
  {Zhevlakov}}]{Kirpichnikov:2020tcf}%
  \BibitemOpen
  \bibfield  {author} {\bibinfo {author} {\bibfnamefont {D.~V.}\ \bibnamefont
  {Kirpichnikov}}, \bibinfo {author} {\bibfnamefont {V.~E.}\ \bibnamefont
  {Lyubovitskij}},\ and\ \bibinfo {author} {\bibfnamefont {A.~S.}\ \bibnamefont
  {Zhevlakov}},\ }\href {https://doi.org/10.1103/PhysRevD.102.095024}
  {\bibfield  {journal} {\bibinfo  {journal} {Phys. Rev. D}\ }\textbf {\bibinfo
  {volume} {102}},\ \bibinfo {pages} {095024} (\bibinfo {year} {2020})},\
  \Eprint {https://arxiv.org/abs/2002.07496} {arXiv:2002.07496 [hep-ph]}
  \BibitemShut {NoStop}%
\bibitem [{\citenamefont {Amaral}\ \emph {et~al.}(2021)\citenamefont {Amaral},
  \citenamefont {Cerde\~no}, \citenamefont {Cheek},\ and\ \citenamefont
  {Foldenauer}}]{Amaral:2021rzw}%
  \BibitemOpen
  \bibfield  {author} {\bibinfo {author} {\bibfnamefont {D.~W.~P.}\
  \bibnamefont {Amaral}}, \bibinfo {author} {\bibfnamefont {D.~G.}\
  \bibnamefont {Cerde\~no}}, \bibinfo {author} {\bibfnamefont {A.}~\bibnamefont
  {Cheek}},\ and\ \bibinfo {author} {\bibfnamefont {P.}~\bibnamefont
  {Foldenauer}},\ }\Eprint {https://arxiv.org/abs/2104.03297} {arXiv:2104.03297
  [hep-ph]}  (\bibinfo {year} {2021})\BibitemShut {NoStop}%
\bibitem [{\citenamefont {Foot}(1991)}]{Foot:1990mn}%
  \BibitemOpen
  \bibfield  {author} {\bibinfo {author} {\bibfnamefont {R.}~\bibnamefont
  {Foot}},\ }\href {https://doi.org/10.1142/S0217732391000543} {\bibfield
  {journal} {\bibinfo  {journal} {Mod. Phys. Lett. A}\ }\textbf {\bibinfo
  {volume} {6}},\ \bibinfo {pages} {527} (\bibinfo {year} {1991})}\BibitemShut
  {NoStop}%
\bibitem [{\citenamefont {He}\ \emph {et~al.}(1991{\natexlab{a}})\citenamefont
  {He}, \citenamefont {Joshi}, \citenamefont {Lew},\ and\ \citenamefont
  {Volkas}}]{PhysRevD.43.R22}%
  \BibitemOpen
  \bibfield  {author} {\bibinfo {author} {\bibfnamefont {X.~G.}\ \bibnamefont
  {He}}, \bibinfo {author} {\bibfnamefont {G.~C.}\ \bibnamefont {Joshi}},
  \bibinfo {author} {\bibfnamefont {H.}~\bibnamefont {Lew}},\ and\ \bibinfo
  {author} {\bibfnamefont {R.~R.}\ \bibnamefont {Volkas}},\ }\href
  {https://doi.org/10.1103/PhysRevD.43.R22} {\bibfield  {journal} {\bibinfo
  {journal} {Phys. Rev. D}\ }\textbf {\bibinfo {volume} {43}},\ \bibinfo
  {pages} {R22} (\bibinfo {year} {1991}{\natexlab{a}})}\BibitemShut {NoStop}%
\bibitem [{\citenamefont {He}\ \emph {et~al.}(1991{\natexlab{b}})\citenamefont
  {He}, \citenamefont {Joshi}, \citenamefont {Lew},\ and\ \citenamefont
  {Volkas}}]{PhysRevD.44.2118}%
  \BibitemOpen
  \bibfield  {author} {\bibinfo {author} {\bibfnamefont {X.-G.}\ \bibnamefont
  {He}}, \bibinfo {author} {\bibfnamefont {G.~C.}\ \bibnamefont {Joshi}},
  \bibinfo {author} {\bibfnamefont {H.}~\bibnamefont {Lew}},\ and\ \bibinfo
  {author} {\bibfnamefont {R.~R.}\ \bibnamefont {Volkas}},\ }\href
  {https://doi.org/10.1103/PhysRevD.44.2118} {\bibfield  {journal} {\bibinfo
  {journal} {Phys. Rev. D}\ }\textbf {\bibinfo {volume} {44}},\ \bibinfo
  {pages} {2118} (\bibinfo {year} {1991}{\natexlab{b}})}\BibitemShut {NoStop}%
\bibitem [{\citenamefont {Baek}\ and\ \citenamefont {Ko}(2009)}]{Baek_2009}%
  \BibitemOpen
  \bibfield  {author} {\bibinfo {author} {\bibfnamefont {S.}~\bibnamefont
  {Baek}}\ and\ \bibinfo {author} {\bibfnamefont {P.}~\bibnamefont {Ko}},\
  }\href {https://doi.org/10.1088/1475-7516/2009/10/011} {\bibfield  {journal}
  {\bibinfo  {journal} {JCAP}\ }\textbf {\bibinfo {volume} {2009}}\bibinfo
  {number} { (10)},\ \bibinfo {pages} {011}}\BibitemShut {NoStop}%
\bibitem [{\citenamefont {Banerjee~{\it et al.}
  [NA64~Collaboration]}(2019)}]{Banerjee:2019na64mu}%
  \BibitemOpen
\bibfield  {number} {  }\bibfield  {author} {\bibinfo {author} {\bibfnamefont
  {D.}~\bibnamefont {Banerjee~{\it et al.} [NA64~Collaboration]}},\ }\Eprint
  {https://arxiv.org/abs/2019-002/ SPSC-P-359} {CERN-SPSC:2019-002/ SPSC-P-359
  [hep-ph]}  (\bibinfo {year} {2019})\BibitemShut {NoStop}%
\bibitem [{\citenamefont {Kahn}\ \emph {et~al.}(2018)\citenamefont {Kahn},
  \citenamefont {Krnjaic}, \citenamefont {Tran},\ and\ \citenamefont
  {Whitbeck}}]{Kahn:2018cqs}%
  \BibitemOpen
  \bibfield  {author} {\bibinfo {author} {\bibfnamefont {Y.}~\bibnamefont
  {Kahn}}, \bibinfo {author} {\bibfnamefont {G.}~\bibnamefont {Krnjaic}},
  \bibinfo {author} {\bibfnamefont {N.}~\bibnamefont {Tran}},\ and\ \bibinfo
  {author} {\bibfnamefont {A.}~\bibnamefont {Whitbeck}},\ }\href
  {https://doi.org/10.1007/JHEP09(2018)153} {\bibfield  {journal} {\bibinfo
  {journal} {JHEP}\ }\textbf {\bibinfo {volume} {09}},\ \bibinfo {pages}
  {153}},\ \Eprint {https://arxiv.org/abs/1804.03144} {arXiv:1804.03144
  [hep-ph]} \BibitemShut {NoStop}%
\bibitem [{\citenamefont {Doble}\ \emph {et~al.}(1994)\citenamefont {Doble},
  \citenamefont {Gatignon}, \citenamefont {von Holtey},\ and\ \citenamefont
  {Novoskoltsev}}]{Doble:1994np}%
  \BibitemOpen
  \bibfield  {author} {\bibinfo {author} {\bibfnamefont {N.}~\bibnamefont
  {Doble}}, \bibinfo {author} {\bibfnamefont {L.}~\bibnamefont {Gatignon}},
  \bibinfo {author} {\bibfnamefont {G.}~\bibnamefont {von Holtey}},\ and\
  \bibinfo {author} {\bibfnamefont {F.}~\bibnamefont {Novoskoltsev}},\ }\href
  {https://doi.org/10.1016/0168-9002(94)90212-7} {\bibfield  {journal}
  {\bibinfo  {journal} {Nucl. Instrum. Meth. A}\ }\textbf {\bibinfo {volume}
  {343}},\ \bibinfo {pages} {351} (\bibinfo {year} {1994})}\BibitemShut
  {NoStop}%
\bibitem [{\citenamefont {Kirpichnikov}\ \emph {et~al.}(2021)\citenamefont
  {Kirpichnikov}, \citenamefont {Sieber}, \citenamefont {Molina~Bueno},
  \citenamefont {Crivelli},\ and\ \citenamefont
  {Kirsanov}}]{kirpichnikov2021probing}%
  \BibitemOpen
  \bibfield  {author} {\bibinfo {author} {\bibfnamefont {D.~V.}\ \bibnamefont
  {Kirpichnikov}}, \bibinfo {author} {\bibfnamefont {H.}~\bibnamefont
  {Sieber}}, \bibinfo {author} {\bibfnamefont {L.}~\bibnamefont
  {Molina~Bueno}}, \bibinfo {author} {\bibfnamefont {P.}~\bibnamefont
  {Crivelli}},\ and\ \bibinfo {author} {\bibfnamefont {M.~M.}\ \bibnamefont
  {Kirsanov}},\ }\href {https://doi.org/10.1103/PhysRevD.104.076012} {\bibfield
   {journal} {\bibinfo  {journal} {Phys. Rev. D}\ }\textbf {\bibinfo {volume}
  {104}},\ \bibinfo {pages} {076012} (\bibinfo {year} {2021})}\BibitemShut
  {NoStop}%
\bibitem [{\citenamefont {Andreas}\ \emph {et~al.}(2013)\citenamefont
  {Andreas}, \citenamefont {Donskov}, \citenamefont {Crivelli}, \citenamefont
  {Gardikiotis}, \citenamefont {Gninenko}, \citenamefont {Golubev},
  \citenamefont {Guber}, \citenamefont {Ivashkin}, \citenamefont {Kirsanov},
  \citenamefont {Krasnikov}, \citenamefont {Matveev}, \citenamefont
  {Mikhailov}, \citenamefont {Musienko}, \citenamefont {Polyakov},
  \citenamefont {Ringwald}, \citenamefont {Rubbia}, \citenamefont {Samoylenko},
  \citenamefont {Semertzidis},\ and\ \citenamefont
  {Zioutas}}]{Andreas:2013lya}%
  \BibitemOpen
  \bibfield  {author} {\bibinfo {author} {\bibfnamefont {S.}~\bibnamefont
  {Andreas}}, \bibinfo {author} {\bibfnamefont {S.~V.}\ \bibnamefont
  {Donskov}}, \bibinfo {author} {\bibfnamefont {P.}~\bibnamefont {Crivelli}},
  \bibinfo {author} {\bibfnamefont {A.}~\bibnamefont {Gardikiotis}}, \bibinfo
  {author} {\bibfnamefont {S.~N.}\ \bibnamefont {Gninenko}}, \bibinfo {author}
  {\bibfnamefont {N.~A.}\ \bibnamefont {Golubev}}, \bibinfo {author}
  {\bibfnamefont {F.~F.}\ \bibnamefont {Guber}}, \bibinfo {author}
  {\bibfnamefont {A.~P.}\ \bibnamefont {Ivashkin}}, \bibinfo {author}
  {\bibfnamefont {M.~M.}\ \bibnamefont {Kirsanov}}, \bibinfo {author}
  {\bibfnamefont {N.~V.}\ \bibnamefont {Krasnikov}}, \bibinfo {author}
  {\bibfnamefont {V.~A.}\ \bibnamefont {Matveev}}, \bibinfo {author}
  {\bibfnamefont {Y.~V.}\ \bibnamefont {Mikhailov}}, \bibinfo {author}
  {\bibfnamefont {Y.~V.}\ \bibnamefont {Musienko}}, \bibinfo {author}
  {\bibfnamefont {V.~A.}\ \bibnamefont {Polyakov}}, \bibinfo {author}
  {\bibfnamefont {A.}~\bibnamefont {Ringwald}}, \bibinfo {author}
  {\bibfnamefont {A.}~\bibnamefont {Rubbia}}, \bibinfo {author} {\bibfnamefont
  {V.~D.}\ \bibnamefont {Samoylenko}}, \bibinfo {author} {\bibfnamefont
  {Y.~K.}\ \bibnamefont {Semertzidis}},\ and\ \bibinfo {author} {\bibfnamefont
  {K.}~\bibnamefont {Zioutas}},\ }\href@noop {} {\bibinfo {title} {Proposal for
  an experiment to search for light dark matter at the sps}} (\bibinfo {year}
  {2013}),\ \Eprint {https://arxiv.org/abs/1312.3309} {arXiv:1312.3309
  [hep-ex]} \BibitemShut {NoStop}%
\bibitem [{\citenamefont {Gninenko}(2014)}]{Gninenko:2013rka}%
  \BibitemOpen
  \bibfield  {author} {\bibinfo {author} {\bibfnamefont {S.~N.}\ \bibnamefont
  {Gninenko}},\ }\href {https://doi.org/10.1103/PhysRevD.89.075008} {\bibfield
  {journal} {\bibinfo  {journal} {Phys. Rev. D}\ }\textbf {\bibinfo {volume}
  {89}},\ \bibinfo {pages} {075008} (\bibinfo {year} {2014})}\BibitemShut
  {NoStop}%
\bibitem [{\citenamefont {Gninenko}\ \emph
  {et~al.}(2019{\natexlab{a}})\citenamefont {Gninenko}, \citenamefont
  {Kirpichnikov}, \citenamefont {Kirsanov},\ and\ \citenamefont
  {Krasnikov}}]{Gninenko:2019qiv}%
  \BibitemOpen
  \bibfield  {author} {\bibinfo {author} {\bibfnamefont {S.~N.}\ \bibnamefont
  {Gninenko}}, \bibinfo {author} {\bibfnamefont {D.~V.}\ \bibnamefont
  {Kirpichnikov}}, \bibinfo {author} {\bibfnamefont {M.~M.}\ \bibnamefont
  {Kirsanov}},\ and\ \bibinfo {author} {\bibfnamefont {N.~V.}\ \bibnamefont
  {Krasnikov}},\ }\href {https://doi.org/10.1016/j.physletb.2019.07.015}
  {\bibfield  {journal} {\bibinfo  {journal} {Phys. Lett. B}\ }\textbf
  {\bibinfo {volume} {796}},\ \bibinfo {pages} {117} (\bibinfo {year}
  {2019}{\natexlab{a}})},\ \Eprint {https://arxiv.org/abs/1903.07899}
  {arXiv:1903.07899 [hep-ph]} \BibitemShut {NoStop}%
\bibitem [{\citenamefont {Gninenko}\ \emph
  {et~al.}(2019{\natexlab{b}})\citenamefont {Gninenko}, \citenamefont
  {Kirpichnikov},\ and\ \citenamefont {Krasnikov}}]{Gninenko:2018ter}%
  \BibitemOpen
  \bibfield  {author} {\bibinfo {author} {\bibfnamefont {S.~N.}\ \bibnamefont
  {Gninenko}}, \bibinfo {author} {\bibfnamefont {D.~V.}\ \bibnamefont
  {Kirpichnikov}},\ and\ \bibinfo {author} {\bibfnamefont {N.~V.}\ \bibnamefont
  {Krasnikov}},\ }\href {https://doi.org/10.1103/PhysRevD.100.035003}
  {\bibfield  {journal} {\bibinfo  {journal} {Phys. Rev. D}\ }\textbf {\bibinfo
  {volume} {100}},\ \bibinfo {pages} {035003} (\bibinfo {year}
  {2019}{\natexlab{b}})},\ \Eprint {https://arxiv.org/abs/1810.06856}
  {arXiv:1810.06856 [hep-ph]} \BibitemShut {NoStop}%
\bibitem [{\citenamefont {Gninenko}\ \emph {et~al.}(2018)\citenamefont
  {Gninenko}, \citenamefont {Kovalenko}, \citenamefont {Kuleshov},
  \citenamefont {Lyubovitskij},\ and\ \citenamefont
  {Zhevlakov}}]{Gninenko:2018num}%
  \BibitemOpen
  \bibfield  {author} {\bibinfo {author} {\bibfnamefont {S.}~\bibnamefont
  {Gninenko}}, \bibinfo {author} {\bibfnamefont {S.}~\bibnamefont {Kovalenko}},
  \bibinfo {author} {\bibfnamefont {S.}~\bibnamefont {Kuleshov}}, \bibinfo
  {author} {\bibfnamefont {V.~E.}\ \bibnamefont {Lyubovitskij}},\ and\ \bibinfo
  {author} {\bibfnamefont {A.~S.}\ \bibnamefont {Zhevlakov}},\ }\href
  {https://doi.org/10.1103/PhysRevD.98.015007} {\bibfield  {journal} {\bibinfo
  {journal} {Phys. Rev. D}\ }\textbf {\bibinfo {volume} {98}},\ \bibinfo
  {pages} {015007} (\bibinfo {year} {2018})},\ \Eprint
  {https://arxiv.org/abs/1804.05550} {arXiv:1804.05550 [hep-ph]} \BibitemShut
  {NoStop}%
\bibitem [{\citenamefont {Bernhard}\ \emph {et~al.}(2020)\citenamefont
  {Bernhard} \emph {et~al.}}]{Bernhard:2019jqz}%
  \BibitemOpen
  \bibfield  {author} {\bibinfo {author} {\bibfnamefont {J.}~\bibnamefont
  {Bernhard}} \emph {et~al.},\ }\href {https://doi.org/10.1063/5.0008957}
  {\bibfield  {journal} {\bibinfo  {journal} {AIP Conf. Proc.}\ }\textbf
  {\bibinfo {volume} {2249}},\ \bibinfo {pages} {030035} (\bibinfo {year}
  {2020})},\ \Eprint {https://arxiv.org/abs/1911.01498} {arXiv:1911.01498
  [hep-ex]} \BibitemShut {NoStop}%
\bibitem [{\citenamefont {Abbon}\ \emph {et~al.}(2007)\citenamefont {Abbon}
  \emph {et~al.}}]{COMPASS:2007rjf}%
  \BibitemOpen
  \bibfield  {author} {\bibinfo {author} {\bibfnamefont {P.}~\bibnamefont
  {Abbon}} \emph {et~al.} (\bibinfo {collaboration} {COMPASS}),\ }\href
  {https://doi.org/10.1016/j.nima.2007.03.026} {\bibfield  {journal} {\bibinfo
  {journal} {Nucl. Instrum. Meth. A}\ }\textbf {\bibinfo {volume} {577}},\
  \bibinfo {pages} {455} (\bibinfo {year} {2007})},\ \Eprint
  {https://arxiv.org/abs/hep-ex/0703049} {arXiv:hep-ex/0703049} \BibitemShut
  {NoStop}%
\bibitem [{\citenamefont {Agostinelli}\ \emph {et~al.}(2003)\citenamefont
  {Agostinelli} \emph {et~al.}}]{Agostinelli:2002hh}%
  \BibitemOpen
  \bibfield  {author} {\bibinfo {author} {\bibfnamefont {S.}~\bibnamefont
  {Agostinelli}} \emph {et~al.} (\bibinfo {collaboration} {GEANT4}),\ }\href
  {https://doi.org/10.1016/S0168-9002(03)01368-8} {\bibfield  {journal}
  {\bibinfo  {journal} {Nucl. Instrum. Meth. A}\ }\textbf {\bibinfo {volume}
  {506}},\ \bibinfo {pages} {250} (\bibinfo {year} {2003})}\BibitemShut
  {NoStop}%
\bibitem [{\citenamefont {Bondi}\ \emph {et~al.}(2021)\citenamefont {Bondi},
  \citenamefont {Celentano}, \citenamefont {Dusaev}, \citenamefont
  {Kirpichnikov}, \citenamefont {Kirsanov}, \citenamefont {Krasnikov},
  \citenamefont {Marsicano},\ and\ \citenamefont
  {Shchukin}}]{Celentano:2021cna}%
  \BibitemOpen
  \bibfield  {author} {\bibinfo {author} {\bibfnamefont {M.}~\bibnamefont
  {Bondi}}, \bibinfo {author} {\bibfnamefont {A.}~\bibnamefont {Celentano}},
  \bibinfo {author} {\bibfnamefont {R.~R.}\ \bibnamefont {Dusaev}}, \bibinfo
  {author} {\bibfnamefont {D.~V.}\ \bibnamefont {Kirpichnikov}}, \bibinfo
  {author} {\bibfnamefont {M.~M.}\ \bibnamefont {Kirsanov}}, \bibinfo {author}
  {\bibfnamefont {N.~V.}\ \bibnamefont {Krasnikov}}, \bibinfo {author}
  {\bibfnamefont {L.}~\bibnamefont {Marsicano}},\ and\ \bibinfo {author}
  {\bibfnamefont {D.}~\bibnamefont {Shchukin}},\ }\href
  {https://doi.org/10.1016/j.cpc.2021.108129} {\bibfield  {journal} {\bibinfo
  {journal} {Comput. Phys. Commun.}\ }\textbf {\bibinfo {volume} {269}},\
  \bibinfo {pages} {108129} (\bibinfo {year} {2021})},\ \Eprint
  {https://arxiv.org/abs/2101.12192} {arXiv:2101.12192 [hep-ph]} \BibitemShut
  {NoStop}%
\bibitem [{\citenamefont {Brown~et al.}(1980)}]{Brown1980}%
  \BibitemOpen
  \bibfield  {author} {\bibinfo {author} {\bibfnamefont {K.}~\bibnamefont
  {Brown~et al.}},\ }\href@noop {} {\bibfield  {journal} {\bibinfo  {journal}
  {CERN Yellow Reports: Monographs}\ } (\bibinfo {year} {1980})}\BibitemShut
  {NoStop}%
\bibitem [{\citenamefont {Iselin}\ and\ \citenamefont
  {Brown}(1974)}]{Turtle1974}%
  \BibitemOpen
  \bibfield  {author} {\bibinfo {author} {\bibfnamefont {C.}~\bibnamefont
  {Iselin}}\ and\ \bibinfo {author} {\bibfnamefont {K.}~\bibnamefont {Brown}},\
  }\href@noop {} {\bibfield  {journal} {\bibinfo  {journal} {CERN Yellow
  Reports: Monographs}\ } (\bibinfo {year} {1974})}\BibitemShut {NoStop}%
\bibitem [{\citenamefont {Iselin}(1974)}]{Iselin1974}%
  \BibitemOpen
  \bibfield  {author} {\bibinfo {author} {\bibfnamefont {C.}~\bibnamefont
  {Iselin}},\ }\href@noop {} {\bibfield  {journal} {\bibinfo  {journal} {CERN
  74-17 Laboratory II}\ } (\bibinfo {year} {1974})}\BibitemShut {NoStop}%
\bibitem [{\citenamefont {Dobbs}\ and\ \citenamefont
  {Hansen}(2000)}]{Dobbs:684090}%
  \BibitemOpen
  \bibfield  {author} {\bibinfo {author} {\bibfnamefont {M.}~\bibnamefont
  {Dobbs}}\ and\ \bibinfo {author} {\bibfnamefont {J.~B.}\ \bibnamefont
  {Hansen}},\ }\href {http://cds.cern.ch/record/684090} {\bibfield  {journal}
  {\bibinfo  {journal} {ATL-SOFT-2000-001}\ } (\bibinfo {year}
  {2000})}\BibitemShut {NoStop}%
\bibitem [{\citenamefont {Rauch}\ and\ \citenamefont
  {Schl\"uter}(2015)}]{Rauch:2014wta}%
  \BibitemOpen
  \bibfield  {author} {\bibinfo {author} {\bibfnamefont {J.}~\bibnamefont
  {Rauch}}\ and\ \bibinfo {author} {\bibfnamefont {T.}~\bibnamefont
  {Schl\"uter}},\ }\href {https://doi.org/10.1088/1742-6596/608/1/012042}
  {\bibfield  {journal} {\bibinfo  {journal} {J. Phys. Conf. Ser.}\ }\textbf
  {\bibinfo {volume} {608}},\ \bibinfo {pages} {012042} (\bibinfo {year}
  {2015})},\ \Eprint {https://arxiv.org/abs/1410.3698} {arXiv:1410.3698
  [physics.ins-det]} \BibitemShut {NoStop}%
\bibitem [{\citenamefont {S.P.~Denisov}\ \emph {et~al.}(1973)\citenamefont
  {S.P.~Denisov}, \citenamefont {V.}, \citenamefont {P.}, \citenamefont {A.I.},
  \citenamefont {D.},\ and\ \citenamefont {A.}}]{DENISOV197362}%
  \BibitemOpen
  \bibfield  {author} {\bibinfo {author} {\bibfnamefont {S.~P.}\ \bibnamefont
  {S.P.~Denisov}}, \bibinfo {author} {\bibfnamefont {D.~S.}\ \bibnamefont
  {V.}}, \bibinfo {author} {\bibfnamefont {G.~Y.}\ \bibnamefont {P.}}, \bibinfo
  {author} {\bibfnamefont {K.~R. N.~P.}\ \bibnamefont {A.I.}}, \bibinfo
  {author} {\bibfnamefont {P.~Y.}\ \bibnamefont {D.}},\ and\ \bibinfo {author}
  {\bibfnamefont {S.~D.}\ \bibnamefont {A.}},\ }\href
  {https://doi.org/https://doi.org/10.1016/0550-3213(73)90351-9} {\bibfield
  {journal} {\bibinfo  {journal} {Nucl. Phys. B}\ }\textbf {\bibinfo {volume}
  {61}},\ \bibinfo {pages} {62} (\bibinfo {year} {1973})}\BibitemShut {NoStop}%
\bibitem [{\citenamefont {Aalste}\ \emph {et~al.}(1993)\citenamefont {Aalste}
  \emph {et~al.}}]{Aalste1993}%
  \BibitemOpen
  \bibfield  {author} {\bibinfo {author} {\bibfnamefont {M.}~\bibnamefont
  {Aalste}, \bibfnamefont {M.~Andlinger}} \emph {et~al.},\ }\href
  {https://doi.org/doi.org/10.1007/BF01650426} {\bibfield  {journal} {\bibinfo
  {journal} {Z. Phys. C}\ }\textbf {\bibinfo {volume} {60}},\ \bibinfo {pages}
  {1} (\bibinfo {year} {1993})}\BibitemShut {NoStop}%
\bibitem [{\citenamefont {Sandler}\ \emph {et~al.}(1990)\citenamefont {Sandler}
  \emph {et~al.}}]{PhysRevD.42.759}%
  \BibitemOpen
  \bibfield  {author} {\bibinfo {author} {\bibfnamefont {P.~H.}\ \bibnamefont
  {Sandler}} \emph {et~al.},\ }\href {https://doi.org/10.1103/PhysRevD.42.759}
  {\bibfield  {journal} {\bibinfo  {journal} {Phys. Rev. D}\ }\textbf {\bibinfo
  {volume} {42}},\ \bibinfo {pages} {759} (\bibinfo {year} {1990})}\BibitemShut
  {NoStop}%
\bibitem [{\citenamefont {Chaudhuri}\ and\ \citenamefont
  {Sinha}(1965)}]{Chaudhuri1965}%
  \BibitemOpen
  \bibfield  {author} {\bibinfo {author} {\bibfnamefont {N.}~\bibnamefont
  {Chaudhuri}}\ and\ \bibinfo {author} {\bibfnamefont {M.~S.}\ \bibnamefont
  {Sinha}},\ }\href {https://doi.org/10.1007/BF02734821} {\bibfield  {journal}
  {\bibinfo  {journal} {Il Nuovo Cimento}\ }\textbf {\bibinfo {volume} {35}},\
  \bibinfo {pages} {13} (\bibinfo {year} {1965})}\BibitemShut {NoStop}%
\bibitem [{\citenamefont {Kelner}\ \emph {et~al.}(1995)\citenamefont {Kelner}
  \emph {et~al.}}]{Kelner95}%
  \BibitemOpen
  \bibfield  {author} {\bibinfo {author} {\bibfnamefont {S.}~\bibnamefont
  {Kelner}} \emph {et~al.},\ }\href@noop {} {\emph {\bibinfo {title} {About
  cross section for high-energy muon bremsstrahlung}}},\ \bibinfo {type}
  {Preprint MEPhI 024-95}\ (\bibinfo  {institution} {MEphI},\ \bibinfo {year}
  {1995})\BibitemShut {NoStop}%
\bibitem [{\citenamefont {Akhiezer}\ and\ \citenamefont
  {Berestestsky}(1965)}]{akhiezer1965}%
  \BibitemOpen
  \bibfield  {author} {\bibinfo {author} {\bibfnamefont {A.}~\bibnamefont
  {Akhiezer}}\ and\ \bibinfo {author} {\bibfnamefont {V.}~\bibnamefont
  {Berestestsky}},\ }\href@noop {} {\emph {\bibinfo {title} {{Quantum
  Electrodynamics}}}}\ (\bibinfo  {publisher} {{Interscience Publishers}},\
  \bibinfo {address} {{Geneva}},\ \bibinfo {year} {1965})\BibitemShut {NoStop}%
\bibitem [{\citenamefont {Bogdanov}\ \emph {et~al.}(2006)\citenamefont
  {Bogdanov}, \citenamefont {Burkhardt}, \citenamefont {Ivanchenko},
  \citenamefont {Kelner}, \citenamefont {Kokoulin}, \citenamefont {Maire},
  \citenamefont {Rybin},\ and\ \citenamefont {Urban}}]{Bogdanov:2006kr}%
  \BibitemOpen
  \bibfield  {author} {\bibinfo {author} {\bibfnamefont {A.~G.}\ \bibnamefont
  {Bogdanov}}, \bibinfo {author} {\bibfnamefont {H.}~\bibnamefont {Burkhardt}},
  \bibinfo {author} {\bibfnamefont {V.~N.}\ \bibnamefont {Ivanchenko}},
  \bibinfo {author} {\bibfnamefont {S.~R.}\ \bibnamefont {Kelner}}, \bibinfo
  {author} {\bibfnamefont {R.~P.}\ \bibnamefont {Kokoulin}}, \bibinfo {author}
  {\bibfnamefont {M.}~\bibnamefont {Maire}}, \bibinfo {author} {\bibfnamefont
  {A.~M.}\ \bibnamefont {Rybin}},\ and\ \bibinfo {author} {\bibfnamefont
  {L.}~\bibnamefont {Urban}},\ }\href {https://doi.org/10.1109/TNS.2006.872633}
  {\bibfield  {journal} {\bibinfo  {journal} {IEEE Trans. Nucl. Sci.}\ }\textbf
  {\bibinfo {volume} {53}},\ \bibinfo {pages} {513} (\bibinfo {year}
  {2006})}\BibitemShut {NoStop}%
\bibitem [{\citenamefont {Mishra}\ \emph {et~al.}(1991)\citenamefont {Mishra}
  \emph {et~al.}}]{PhysRevLett.66.3117}%
  \BibitemOpen
  \bibfield  {author} {\bibinfo {author} {\bibfnamefont {S.~R.}\ \bibnamefont
  {Mishra}} \emph {et~al.} (\bibinfo {collaboration} {CCFR Collaboration}),\
  }\href {https://doi.org/10.1103/PhysRevLett.66.3117} {\bibfield  {journal}
  {\bibinfo  {journal} {Phys. Rev. Lett.}\ }\textbf {\bibinfo {volume} {66}},\
  \bibinfo {pages} {3117} (\bibinfo {year} {1991})}\BibitemShut {NoStop}%
\bibitem [{\citenamefont {Altmannshofer}\ \emph {et~al.}(2014)\citenamefont
  {Altmannshofer}, \citenamefont {Gori}, \citenamefont {Pospelov},\ and\
  \citenamefont {Yavin}}]{PhysRevLett.113.091801}%
  \BibitemOpen
  \bibfield  {author} {\bibinfo {author} {\bibfnamefont {W.}~\bibnamefont
  {Altmannshofer}}, \bibinfo {author} {\bibfnamefont {S.}~\bibnamefont {Gori}},
  \bibinfo {author} {\bibfnamefont {M.}~\bibnamefont {Pospelov}},\ and\
  \bibinfo {author} {\bibfnamefont {I.}~\bibnamefont {Yavin}},\ }\href
  {https://doi.org/10.1103/PhysRevLett.113.091801} {\bibfield  {journal}
  {\bibinfo  {journal} {Phys. Rev. Lett.}\ }\textbf {\bibinfo {volume} {113}},\
  \bibinfo {pages} {091801} (\bibinfo {year} {2014})}\BibitemShut {NoStop}%
\bibitem [{\citenamefont {Kaneta}\ and\ \citenamefont
  {Shimomura}(2017)}]{10.1093/ptep/ptx050}%
  \BibitemOpen
  \bibfield  {author} {\bibinfo {author} {\bibfnamefont {Y.}~\bibnamefont
  {Kaneta}}\ and\ \bibinfo {author} {\bibfnamefont {T.}~\bibnamefont
  {Shimomura}},\ }\href {https://doi.org/10.1093/ptep/ptx050} {\bibfield
  {journal} {\bibinfo  {journal} {Prog. Theor. Exp. Phys.}\ }\textbf {\bibinfo
  {volume} {2017}} (\bibinfo {year} {2017})},\ \bibinfo {note}
  {053B04}\BibitemShut {NoStop}%
\bibitem [{\citenamefont {Lees}\ \emph {et~al.}(2016)\citenamefont {Lees} \emph
  {et~al.}}]{PhysRevD.94.011102}%
  \BibitemOpen
  \bibfield  {author} {\bibinfo {author} {\bibfnamefont {J.~P.}\ \bibnamefont
  {Lees}} \emph {et~al.} (\bibinfo {collaboration} {BaBar Collaboration}),\
  }\href {https://doi.org/10.1103/PhysRevD.94.011102} {\bibfield  {journal}
  {\bibinfo  {journal} {Phys. Rev. D}\ }\textbf {\bibinfo {volume} {94}},\
  \bibinfo {pages} {011102} (\bibinfo {year} {2016})}\BibitemShut {NoStop}%
\bibitem [{\citenamefont {Adachi}\ \emph {et~al.}(2020)\citenamefont {Adachi}
  \emph {et~al.}}]{Belle-II:2019qfb}%
  \BibitemOpen
  \bibfield  {author} {\bibinfo {author} {\bibfnamefont {I.}~\bibnamefont
  {Adachi}} \emph {et~al.} (\bibinfo {collaboration} {Belle-II}),\ }\href
  {https://doi.org/10.1103/PhysRevLett.124.141801} {\bibfield  {journal}
  {\bibinfo  {journal} {Phys. Rev. Lett.}\ }\textbf {\bibinfo {volume} {124}},\
  \bibinfo {pages} {141801} (\bibinfo {year} {2020})},\ \Eprint
  {https://arxiv.org/abs/1912.11276} {arXiv:1912.11276 [hep-ex]} \BibitemShut
  {NoStop}%
\bibitem [{\citenamefont {Ballett}\ \emph {et~al.}(2019)\citenamefont
  {Ballett}, \citenamefont {Hostert}, \citenamefont {Pascoli}, \citenamefont
  {Perez-Gonzalez}, \citenamefont {Tabrizi},\ and\ \citenamefont
  {Funchal}}]{PhysRevD.100.055012}%
  \BibitemOpen
  \bibfield  {author} {\bibinfo {author} {\bibfnamefont {P.}~\bibnamefont
  {Ballett}}, \bibinfo {author} {\bibfnamefont {M.}~\bibnamefont {Hostert}},
  \bibinfo {author} {\bibfnamefont {S.}~\bibnamefont {Pascoli}}, \bibinfo
  {author} {\bibfnamefont {Y.~F.}\ \bibnamefont {Perez-Gonzalez}}, \bibinfo
  {author} {\bibfnamefont {Z.}~\bibnamefont {Tabrizi}},\ and\ \bibinfo {author}
  {\bibfnamefont {R.~Z.}\ \bibnamefont {Funchal}},\ }\href
  {https://doi.org/10.1103/PhysRevD.100.055012} {\bibfield  {journal} {\bibinfo
   {journal} {Phys. Rev. D}\ }\textbf {\bibinfo {volume} {100}},\ \bibinfo
  {pages} {055012} (\bibinfo {year} {2019})}\BibitemShut {NoStop}%
\bibitem [{\citenamefont {Altmannshofer}\ \emph {et~al.}(2019)\citenamefont
  {Altmannshofer}, \citenamefont {Gori}, \citenamefont {Mart\'{\i}n-Albo},
  \citenamefont {Sousa},\ and\ \citenamefont {Wallbank}}]{PhysRevD.100.115029}%
  \BibitemOpen
  \bibfield  {author} {\bibinfo {author} {\bibfnamefont {W.}~\bibnamefont
  {Altmannshofer}}, \bibinfo {author} {\bibfnamefont {S.}~\bibnamefont {Gori}},
  \bibinfo {author} {\bibfnamefont {J.}~\bibnamefont {Mart\'{\i}n-Albo}},
  \bibinfo {author} {\bibfnamefont {A.}~\bibnamefont {Sousa}},\ and\ \bibinfo
  {author} {\bibfnamefont {M.}~\bibnamefont {Wallbank}},\ }\href
  {https://doi.org/10.1103/PhysRevD.100.115029} {\bibfield  {journal} {\bibinfo
   {journal} {Phys. Rev. D}\ }\textbf {\bibinfo {volume} {100}},\ \bibinfo
  {pages} {115029} (\bibinfo {year} {2019})}\BibitemShut {NoStop}%
\bibitem [{\citenamefont {Campajola}\ \emph {et~al.}(2021)\citenamefont
  {Campajola} \emph {et~al.}}]{BelleII:Campajola2021}%
  \BibitemOpen
  \bibfield  {author} {\bibinfo {author} {\bibfnamefont {M.}~\bibnamefont
  {Campajola}} \emph {et~al.} (\bibinfo {collaboration} {Belle-II
  Collaboration}),\ }\href {https://doi.org/10.1088/1402-4896/abfef21}
  {\bibfield  {journal} {\bibinfo  {journal} {Phys. Scr.}\ }\textbf {\bibinfo
  {volume} {96}},\ \bibinfo {pages} {084005} (\bibinfo {year}
  {2021})}\BibitemShut {NoStop}%
\bibitem [{\citenamefont {Kamada}\ and\ \citenamefont
  {Yu}(2015)}]{PhysRevD.92.113004}%
  \BibitemOpen
  \bibfield  {author} {\bibinfo {author} {\bibfnamefont {A.}~\bibnamefont
  {Kamada}}\ and\ \bibinfo {author} {\bibfnamefont {H.-B.}\ \bibnamefont
  {Yu}},\ }\href {https://doi.org/10.1103/PhysRevD.92.113004} {\bibfield
  {journal} {\bibinfo  {journal} {Phys. Rev. D}\ }\textbf {\bibinfo {volume}
  {92}},\ \bibinfo {pages} {113004} (\bibinfo {year} {2015})}\BibitemShut
  {NoStop}%
\bibitem [{\citenamefont {Ahlgren}\ \emph {et~al.}(2013)\citenamefont
  {Ahlgren}, \citenamefont {Ohlsson},\ and\ \citenamefont
  {Zhou}}]{PhysRevLett.111.199001}%
  \BibitemOpen
  \bibfield  {author} {\bibinfo {author} {\bibfnamefont {B.}~\bibnamefont
  {Ahlgren}}, \bibinfo {author} {\bibfnamefont {T.}~\bibnamefont {Ohlsson}},\
  and\ \bibinfo {author} {\bibfnamefont {S.}~\bibnamefont {Zhou}},\ }\href
  {https://doi.org/10.1103/PhysRevLett.111.199001} {\bibfield  {journal}
  {\bibinfo  {journal} {Phys. Rev. Lett.}\ }\textbf {\bibinfo {volume} {111}},\
  \bibinfo {pages} {199001} (\bibinfo {year} {2013})}\BibitemShut {NoStop}%
\bibitem [{\citenamefont {Escudero}\ \emph {et~al.}(2019)\citenamefont
  {Escudero}, \citenamefont {Hooper}, \citenamefont {Krnjaic},\ and\
  \citenamefont {Pierre}}]{Escudero:2019gzq}%
  \BibitemOpen
  \bibfield  {author} {\bibinfo {author} {\bibfnamefont {M.}~\bibnamefont
  {Escudero}}, \bibinfo {author} {\bibfnamefont {D.}~\bibnamefont {Hooper}},
  \bibinfo {author} {\bibfnamefont {G.}~\bibnamefont {Krnjaic}},\ and\ \bibinfo
  {author} {\bibfnamefont {M.}~\bibnamefont {Pierre}},\ }\href
  {https://doi.org/10.1007/JHEP03(2019)071} {\bibfield  {journal} {\bibinfo
  {journal} {JHEP}\ }\textbf {\bibinfo {volume} {03}},\ \bibinfo {pages}
  {071}},\ \Eprint {https://arxiv.org/abs/1901.02010} {arXiv:1901.02010
  [hep-ph]} \BibitemShut {NoStop}%
\end{thebibliography}%

\end{document}